\documentclass[aps,prb,showpacs,citeautoscript,twocolumn,superscriptaddress]{revtex4}

\usepackage{graphicx}
\usepackage{amsmath}
\usepackage{amssymb}

\newcommand{\bea}{\begin{eqnarray*}}
\newcommand{\eea}{\end{eqnarray*}}
\newcommand{\bne}{\begin{equation*}}
\newcommand{\ede}{\end{equation*}}

\newcommand{\bnen}{\begin{equation}}
\newcommand{\eden}{\end{equation}}
\newcommand{\bean}{\begin{eqnarray}}
\newcommand{\eean}{\end{eqnarray}}
\newcommand{\bnsn}{\begin{subequations}}
\newcommand{\edsn}{\end{subequations}}

\newcommand{\bna}{\begin{array}}
\newcommand{\eda}{\end{array}}
\newcommand{\bnm}{\begin{enumerate}}
\newcommand{\edm}{\end{enumerate}}
\newcommand{\bni}{\begin{itemize}}
\newcommand{\edi}{\end{itemize}}

\newcommand{\spinor}[2]{\left(\bna{c} #1 \\[1.5ex] #2 \eda\right)}

\renewcommand{\vec}[1]{\text{\boldmath{$ #1 $}}}

\newcommand{\ku}{K \! \uparrow}
\newcommand{\kd}{K \! \downarrow}
\newcommand{\qu}{K' \! \uparrow}
\newcommand{\qd}{K' \! \downarrow}

\newcommand{\microev} {\rm{\mu eV}}

\newcommand{\ket}[1]{| #1 \rangle}
\newcommand{\sket}[1]{| #1 \rangle_s}

\begin{document}
\title{Spin-valley blockade in carbon nanotube double quantum dots}

\author{Andr\'as P\'alyi}
\affiliation{Department of Physics, University of Konstanz, D-78457 Konstanz, Germany}
\affiliation{Department of Materials Physics, E\"otv\"os University--Budapest, 
P.O. Box 32, H-1517 Budapest, Hungary}

\author{Guido Burkard}
\affiliation{Department of Physics, University of Konstanz, D-78457 Konstanz, Germany}

\date{\today}

\newcommand{\Dso}{\Delta_{\rm SO}}
\begin{abstract}
We present a theoretical study of the Pauli or spin-valley blockade 
for double quantum dots in semiconducting carbon 
nanotubes. 
In our model, we take into account the following characteristic
features of carbon nanotubes: 
(i) fourfold (spin and valley) degeneracy of the quantum-dot levels
(ii) the intrinsic spin-orbit interaction which is enhanced by the tube curvature, and
(iii) valley mixing due to short-range disorder, i.e., substitutional atoms, adatoms, etc.
We find that 
the spin-valley blockade can be lifted in the presence of
short-range disorder, which induces two independent 
random (in magnitude and direction) valley-Zeeman fields in the two dots,
and hence acts similarly to hyperfine interaction in conventional semiconductor
quantum dots.
In the case of strong spin-orbit interaction, we identify a 
parameter regime where the current as the function of an applied axial
magnetic field shows
a zero-field dip with a width controlled by the interdot tunneling amplitude, 
in agreement with recent experiments.
\end{abstract}


\maketitle


\section{Introduction}

Recent developments of experimental techniques allow for 
preparation, manipulation and readout of few-electron spin states in
quantum dots (QDs),\cite{Hanson-rmp}
indicating the strong potential of these systems for future application
in quantum information processing\cite{Loss-divincenzo}. 
A major factor limiting the performance of quantum-dot spin qubits in widely used
III-V semiconductors (e.g., GaAs) is
spin decoherence due to hyperfine interaction with nuclear spins. 
A strategy to suppress spin decoherence is to 
use QDs dominantly 
consisting of nuclear-spin-free isotopes of group IV materials. 
Carbon structures, such as carbon nanotubes (CNTs) or graphene,
are prime candidates for that purpose as the natural abundance 
of spin-carrying $^{13}$C nuclei is very small (1\%). 
This observation has motivated intensive theoretical 
investigation \cite{Trauzettel-spinqubitsingraphene,Recher-grapheneqds,Bulaev-socincntdots,Fischer-prb-cnt,Roy-coulombincnt,Secchi-coulombincnt,Wunsch-coulombincnt,Palyi-spinblockade,Struck-spinrelaxation,Recher-cm-grapheneqdsreview,Rudner-deflectioncoupling,Schnez-graphenebilliard} and
the experimental realization of QDs in carbon nanostructures \cite{Kong-cntqd,Tans-cntqw,
Mason-cntdqds,Graber-cntdoubledots,Kuemmeth-spinorbit-in-cnts,Buitelaar-spinblockade,Churchill-13cntprl,Churchill-cntspinblockade,
Steele-cntdqd,Ponomarenko-diracbilliard,Schnez-graphenebilliard,Stampfer-grapheneqd,Molitor-graphenedoubledot,Liu-prb-coulombblockade,Kuemmeth-review,Chorley-cntspinblockade}.
Further perspectives of carbon-based quantum information processing 
have been opened by proposals suggesting to utilize the valley degree of freedom of
the delocalized electrons as a qubit \cite{Rycerz-valleytronics,Recher-graphenering},
and to exploit the interplay of spin-orbit interaction, valley mixing,
and the bending of CNTs for implementing qubit 
operations\cite{Flensberg-bentnanotubes}.

The Pauli blockade or spin blockade effect
\cite{Hanson-rmp,Ono-spinblockade} in conventional semiconductor 
double QDs (DQDs) has provided a distinct probe of spin physics in
these devices and has been utilized in the past decade 
for various purposes in the context of spin qubits.
A basic application is spin-state initialization and readout in experiments
realizing resonant manipulation of single spins\cite{Koppens-esr,Nowack-esr,Laird-prl-edsr}.
Pulsed-gate techniques combined with the spin blockade setup 
have been used\cite{Petta-science,Petta-landauzener,Ribeiro-stplus}
in qubit manipulation experiments where the information was encoded in the
two-electron spin states $S$ and $T_0$ or $S$ and $T_+$.
Similar experiments have been utilized to prepare the state of the 
nuclear-spin ensemble of the crystal lattice, with the aim of prolonging the 
decoherence
time of the qubit\cite{Reilly-science,Foletti-natphys,Ribeiro-landauzener}.
Furthermore, spin blockade has been proven an efficient tool to gain information
about the mechanisms of spin relaxation and
decoherence, and the corresponding energy scales.
In particular, it has been applied to measure the energy scales of hyperfine\cite{Koppens-spinblockade,Jouravlev-spinblockade}
and spin-orbit interactions\cite{NadjPerge-spinblockade,Danon-spinblockade}.
The implementation of this range of functionalities in carbon-based quantum dots, 
potentially showing improved qubit performance, is an intense ongoing effort\cite{Buitelaar-spinblockade,Churchill-cntspinblockade,Churchill-13cntprl,Chorley-cntspinblockade}.

\begin{figure}
\includegraphics[scale=0.23]{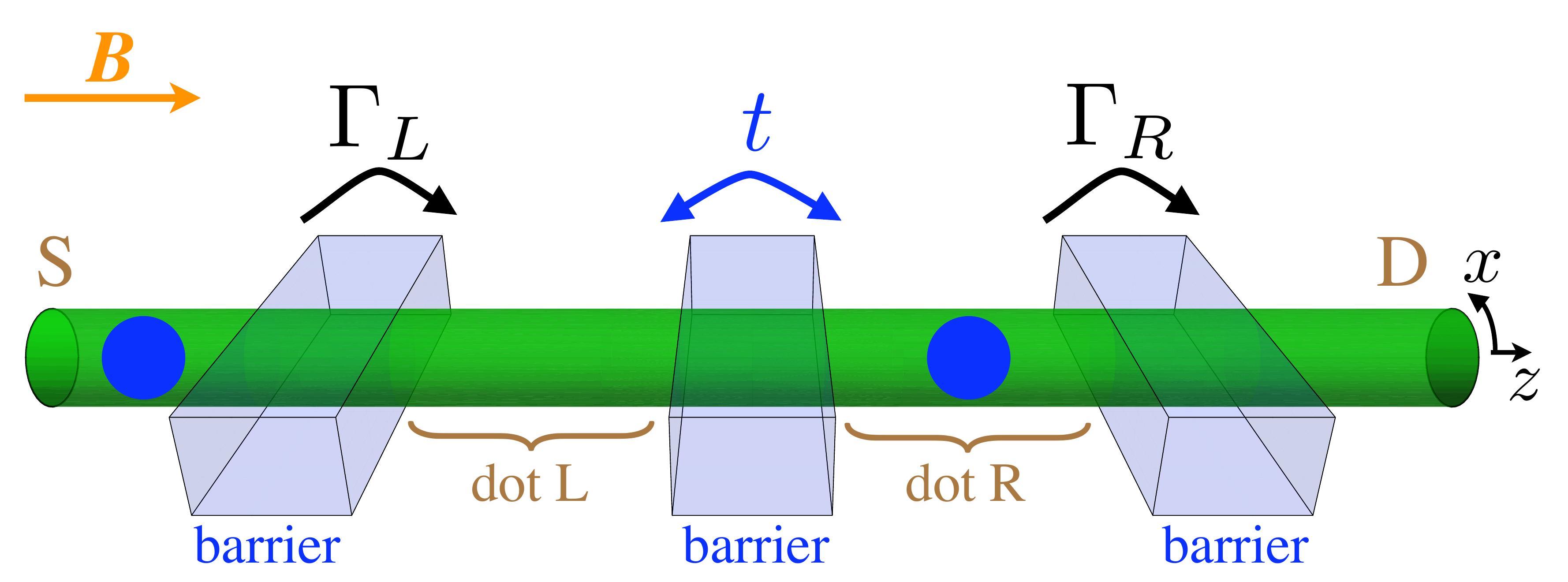}
\caption{\label{fig:cntdqd}
(Color online)
Schematic of the spin-valley blockade setup with a carbon nanotube double quantum 
dot and
an external magnetic field $\vec B$ aligned with the tube axis.
In this regime electrons are transported 
from source (S) to drain (D) while the DQD occupancy changes between single and 
double. 
Spots represent electrons;
the figure shows the (0,1) charge configuration of the double dot.
Lead-dot tunneling rates $\Gamma_L$, $\Gamma_R$ and interdot tunneling
amplitude $t$ are indicated.
}
\end{figure}

In this work, we consider Pauli blockade in a transport setup\cite{Hanson-rmp,Ono-spinblockade}, where 
electrons are transmitted from the source to the drain in a serially coupled DQD
via the $(0,1) \to (1,1) \to (0,2) \to (0,1)$ cycle
(Fig. \ref{fig:cntdqd}).
Here $(n_L,n_R)$ denotes the charge state with $n_L$ ($n_R$) electrons in 
the left (right) QD.
In conventional semiconductor DQDs, 
if the (1,1) and (0,2) states are aligned in energy, then states sharing the same
spin state become hybridized due to interdot tunneling.
The only energetically available (0,2) state has a singlet spin state,
therefore it hybridizes with the (1,1) singlet only, leaving the three (1,1) triplet states without
a (0,2) component. 
This implies that whenever a (1,1) triplet state is occupied in the transport process 
the current is blocked since the (1,1) triplet state cannot decay to a (0,1) state:
the occupation of (1,0) states is energetically forbidden and the source
connected to dot L cannot absorb the electron.
The blockade can be lifted by various mechanisms influencing spin dynamics,
e.g., hyperfine interaction, spin relaxation, etc., resulting in a nonzero
steady-state current, termed as the ``leakage current'' \cite{Hanson-rmp},
through the DQD.

This explanation is altered in the case of electrostatically defined CNT DQDs 
(Fig. \ref{fig:cntdqd}) where the
valley degeneracy of the electronic spectrum is maintained\cite{Churchill-13cntprl,Palyi-spinblockade} ,
resulting in fourfold-degenerate (spin and valley) quantum-dot energy levels
(see Fig. \ref{fig:singledot}).
In this case the six  (0,2) states have \emph{combined spin-valley wave 
functions} which are antisymmetric under particle exchange.
We refer to such states as \emph{supersinglets}. 
The 16-dimensional (1,1) subspace can be separated to a 
six-dimensional supersinglet
subspace and a ten-dimensional subspace of combined spin-valley
wave functions being symmetric under particle exchange, i.e., \emph{supertriplets}.
Hybridization occurs between the (0,2) states and the (1,1) supersinglets,
and the (1,1) supertriplets do not acquire any (0,2) components.
As discussed above for the case of conventional semiconductor DQDs, this 
leads to a blockade of the transport.
The blockade can be lifted by various mechanisms affecting the spin and valley
dynamics. 
To distinguish the cases of conventional and CNT DQDs, we refer to them as spin 
blockade and spin-valley blockade in the following, respectively.

Here we focus on recent
experiments\cite{Churchill-cntspinblockade,Churchill-13cntprl}
observing the spin-valley blockade in clean
CNT DQDs with natural (1\%) and enriched (99\%) $^{13}$C abundance.
Charge sensing data\cite{Churchill-13cntprl} indicates that in these samples the
valley degeneracy was maintained in contrast to other 
observations\cite{Buitelaar-spinblockade,Chorley-cntspinblockade}.
In the case of the isotope-enriched samples, a zero-field peak has been observed\cite{Churchill-cntspinblockade} in
the magnetic-field dependence of the leakage current at small interdot tunneling.
Following a model developed for GaAs DQDs\cite{Jouravlev-spinblockade}, 
this feature has been attributed to hyperfine interaction, although the corresponding
energy scale inferred from the measurement is two orders of magnitude larger than
the theoretical estimates\cite{Yazyev-graphenehf,Fischer-prb-cnt}. 
This discrepancy has not yet been explained.
At large interdot tunneling, the observed magnetotransport data show
a zero-field dip\cite{Churchill-cntspinblockade,Churchill-13cntprl}
with a width controlled by the transparency of the interdot tunneling barrier,
irrespective of the dominant isotope species.
In InAs DQDs, a similar feature has been measured 
recently\cite{Pfund-inasprl,NadjPerge-spinblockade}, and good agreement 
has been found with a phenomenological model incorporating spin-orbit-enabled
spin-flip interdot tunneling and spin relaxation\cite{Danon-spinblockade}. 
The same mechanism might be responsible for the observed magnetotransport
in CNT DQDs as well.

In this work, we provide an alternative explanation of the zero-field
dip found in the magnetotransport curve of CNT DQDs in the spin-valley blockade
regime.
Using a microscopic model, we argue that the disorder-induced valley dynamics is
different in the two dots, resulting in the lifting of the spin-valley blockade
and allowing for a finite current through the DQD.
In this mechanism, disorder plays a role analogous to hyperfine interaction in 
conventional semiconductor DQDs. 

To show this, we set up a model Hamiltonian for the DQD accounting for 
the following unconventional properties of CNT QDs: 
(i) fourfold (spin and valley) degeneracy of the QD energy levels,
(ii) the intrinsic spin-orbit interaction which is enhanced by the tube curvature and
induces an energy splitting between Kramers pairs, and
(iii) valley mixing due to short-range disorder, i.e., substitutional atoms, adatoms, etc.
We provide a microscopic analysis of property (iii), resulting in an effective Hamiltonian
for a single fourfold degenerate QD level.
We find that disorder appears in this Hamiltonian as a random 
(in magnitude and direction) effective magnetic
field acting on the valley degree of freedom.
We express this valley-Zeeman field 
as a function of the disorder configuration and the envelope function 
of the electron occupying the QD.
Our transport calculations are based on a Born-Markov master equation.
The main finding of this work is that the disorder-induced valley-Zeeman fields 
provide a mechanism that lifts the spin-valley blockade.
Depending on the relative significance of spin-orbit interaction and disorder, 
we identify different patterns in the magnetic-field dependence of the 
steady-state current.
In the case of strong spin-orbit interaction, we find a zero-field dip in the 
magnetotransport curve, in agreement with recent experiments,
however our model does not include spin-flip tunneling or spin-relaxation processes.
In the case of strong disorder, we find that the magnetotransport curve can show
both a zero-field dip and peak, depending on the disorder configuration.

The rest of the paper is organized as follows.
In Sec. \ref{sec:heff}, we provide a microscopic analysis of short-range 
disorder in a CNT QD. 
In Sec. \ref{sec:transportmodel}, the model Hamiltonian of the CNT DQD and the
master-equation approach is described.
In Secs. \ref{sec:strongsoi} and \ref{sec:strongdisorder},
we study the magnetotransport in the spin-valley blockade regime in the cases
of strong spin-orbit interaction and strong disorder, respectively.
Our conclusions are presented in Sec. \ref{sec:discussion}.


\section{Short-range disorder in the quantum dot Hamiltonian}
\label{sec:heff}

\begin{figure}
\includegraphics[scale=0.4]{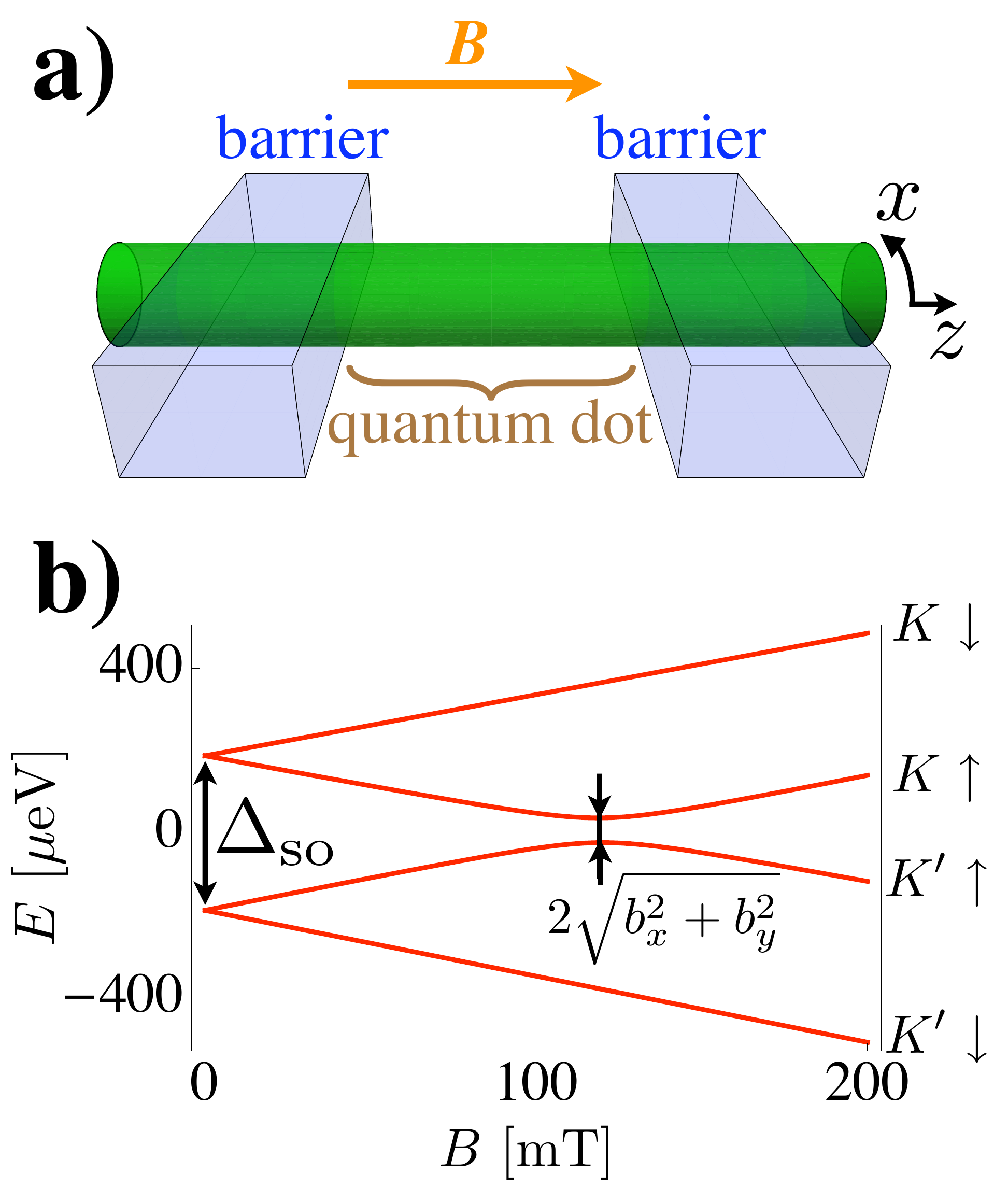}
\caption{\label{fig:singledot}
(Color online)
(a) Schematic of a quantum dot in a carbon nanotube, 
with an external magnetic field $\vec B$ aligned with the tube axis.
(b) Magnetic-field dependence of the spin-orbit-split single-electron ground state 
sublevels of a 
nanotube quantum dot, obtained from diagonalizing $H_0+H_{\rm eff,dis}$ 
[see Eqs. \eqref{eq:srd} and \eqref{eq:disorderindependent}].
Spin and valley quantum numbers of the energy levels are indicated on the right.}
\end{figure}

In this section, we consider a single electrostatically defined QD in a 
semiconducting CNT, in the presence of a homogeneous magnetic field.
Our aim is to derive a $4\times 4$ effective Hamiltonian describing the effect of 
the short-range disorder present in the CNT on a fourfold (spin and valley)
degenerate state of the QD. 
We show that in this effective Hamiltonian the short-range disorder appears as an
effective magnetic (Zeeman) field acting on the valley degree of freedom,
and having a random magnitude and direction.

First we consider a CNT QD model without spin-orbit interaction and short-range
disorder.
We choose the $z$ axis of the coordinate system as aligned with the 
axis of the CNT, and the $x$ coordinate is measured along the circumference of the nanotube as indicated in Fig. \ref{fig:singledot}a.
 
The four tight-binding wave functions, corresponding to a fourfold degenerate 
single-particle energy level of the QD have the form\cite{ando-aob1}
\bnen
(\psi_{vs})_{l \sigma} \equiv 
(\psi_{v})_{l\sigma} \chi_s =
\sqrt{\Omega_{\rm cell}} e^{i (v\vec K\cdot \vec r_{l \sigma} + \varphi_{v\sigma})}
\Psi_\sigma^{(v)}(\vec r_{l \sigma}) \chi_s,
\eden
where 
$s\in(\uparrow,\downarrow) \equiv (+,-)$
and 
$v\in(K,K') \equiv (+,-)$ are
spin and valley quantum numbers.
Furthermore, $\sigma\in\{A,B\}$ is the sublattice index, $l$ is the unit-cell index,
 $\Omega_{\rm cell}$ is the unit-cell area,
 $\vec r_{l \sigma}=(x_{l\sigma},z_{l\sigma})$ 
is the position of the carbon atom on sublattice $\sigma$ in the $l$th unit cell,
the phase factors\cite{ando-aob1} in the exponential are 
$\varphi_{K,A} = \varphi_{K',B} = 0$,
$\varphi_{K',A} = \eta$, and $\varphi_{K,B}=\eta-\pi/3$ with
$\eta$ being the chiral angle of the CNT,
and $\chi_+=(1,0)$ and $\chi_-=(0,1)$ 
are the two possible spin states with axial polarization. 
The four smoothly varying envelope functions $\Psi_\sigma^{(v)}$ can be obtained by
solving the Dirac-type envelope function equations\cite{ando-aob1}
for $v\in(+,-)$,
\bnen
\left[v_F (\sigma_x p_x + v \sigma_y p_z)  + V_{\rm conf}(z) \right]
\spinor{\Psi^{(v)}_A}{\Psi^{(v)}_B}
= E \spinor{\Psi^{(v)}_A}{\Psi^{(v)}_B}.
\eden
Here $\sigma_x$ and $\sigma_y$ are Pauli matrices,
corresponding to the sublattice degree of freedom and
$V_{\rm conf}(z)$ is a smooth confinement potential, e.g., 
induced by electrostatic gates.
Note that our choice of the coordinate system [see Fig. \ref{fig:singledot}(a)] implies
that $p_z$ (and not $p_y$) appears in the envelope Hamiltonian.
The functions $\Psi_\sigma^{(v)}$ and $\psi_{vs}$ are normalized,
\begin{subequations}
 \bean
   1&=& \int_0^{2\pi R} dx\ 
            \int_{-\infty}^{\infty }dz\  
            \left( |\Psi_A^{(v)}(\vec r)|^2+|\Psi_B^{(v)}(\vec r)|^2 \right),\\
   1 &=& \sum_{l\sigma}(\psi_{vs})_{l\sigma}^\dag (\psi_{vs})_{l\sigma},
 \eean
\end{subequations}
where $R$ is the radius of the nanotube.

\newcommand{\efftau}{\tilde \tau}
\newcommand{\efftauz}{\efftau_z}
Our goal is to set up a $4\times 4$ effective Hamiltonian
describing the valley mixing due to short-range disorder. 
Short-range disorder can be caused by any kind of atomic faults of the crystalline
structure: substitutional or interstitial atoms, vacancies, adatoms, etc.
We take into account short-range disorder in the tight-binding model 
 as a static random on-site potential $V_i$,
 i.e., $\left(H_{\rm dis,TB}\right)_{i,j} = V_i \delta_{ij}$.
 [$i=(l\sigma)$ is an index combining the unit cell index
 $l$ and the sublattice index $\sigma$.]
Without the loss of generality, we can assume that the disorder potential has
zero mean, $\langle V_i \rangle = 0$.
The short-range impurities are typically charge neutral, and therefore the 
interaction between them is weak.
This suggests that the random on-site potential
is spatially uncorrelated, 
$\langle V_i V_j \rangle =  \delta_{ij} \langle V_i^2 \rangle$.
A further plausible assumption is that the CNT is homogeneous.
Motivated by these observations,  
we model the disorder potential on the different sites
as independent and identically distributed random variables.
Since we focus on valley effects, we neglect possible sources of 
spin-dependent short-range disorder, such as 
hyperfine interaction due to $^{13}$C 
atoms\cite{Palyi-spinblockade} and
adatom-enhanced spin-orbit interaction\cite{CastroNeto-graphanesoi},
for example.

To derive an effective $4\times 4$ Hamiltonian describing the effect
of the short-range disorder, we project the tight-binding disorder Hamiltonian 
$H_{\rm dis,TB}$ onto the four-dimensional subspace of interest.
The corresponding projector is 
\bnen
P=\sum_{vs} |\psi_{vs}\rangle\langle\psi_{vs}|.
\eden
The same method has been used recently by us to analyze the effect of hyperfine
interaction in carbon-based QDs\cite{Palyi-spinblockade}.
The obtained effective Hamiltonian is 
\bean
H_{\rm dis,eff} &=& 
P H_{\rm dis,TB} P \nonumber \\
&=& 
(b_0 \efftau_0 +
b_x \efftau_x +
b_y \efftau_y +
b_z \efftau_z)\otimes s_0 \nonumber \\
&\equiv& 
(b_0 \efftau_0 + \vec b \cdot \vec \efftau) \otimes s_0,
\label{eq:srd}
\eean
where
\bnen
\label{eq:valleyfields}
   b_k = \Omega_{\rm cell} \sum_{l\sigma} V_{l\sigma} 
                        F^{(k)}_{l\sigma}
\eden
for $k \in \{0,x,y,z\}$.
Here,
$F^{(0)}_{l\sigma} = \sum_{v}|\Psi_\sigma^{(v)}(\vec r_{l\sigma})|^2/2$, 
$F^{(z)}_{l\sigma} = \sum_{v}v|\Psi_\sigma^{(v)}(\vec r_{l\sigma})|^2/2$, and
$F^{(x/y)}_{l\sigma} = {\rm Re/Im}\left(e^{2i\vec K \cdot \vec r_{l\sigma}} 
e^{i(\varphi_{+\sigma} - \varphi_{-\sigma})}
\Psi_\sigma^{(-)*} (\vec r_{l\sigma}) \Psi_\sigma^{(+)} (\vec r_{l\sigma})
\right)$.
The operators $\efftau_{0,x,y,z}$
are natural representations
of the Pauli matrices on the two-dimensional Hilbert space spanned by
$\psi_K$ and $\psi_{K'}$, i.e.,
\begin{subequations}
 \label{eq:efftau}
 \bean
  \efftau_{0} &=&
  |\psi_K \rangle \langle \psi_{K}| + |\psi_{K'} \rangle \langle \psi_{K'}|, \\
  \efftau_{x} &=& 
  |\psi_K \rangle \langle \psi_{K'}| + |\psi_{K'} \rangle \langle \psi_{K}|, \\ 
  \efftau_{y} &=& 
  -i |\psi_{K} \rangle \langle \psi_{K'}| +i |\psi_{K'} \rangle \langle \psi_{K}|,\\
  \efftau_{z} &=&
  |\psi_K \rangle \langle \psi_{K}| - |\psi_{K'} \rangle \langle \psi_{K'}|,
 \eean
\end{subequations}
and $s_0$ is the unit matrix in spin space.
The first term in Eq. \eqref{eq:srd}, proportional to $\efftau_0$,
is just a simultaneous shift of the four energy levels.
The second term in Eq. \eqref{eq:srd}
is reminiscent of a Zeeman coupling Hamiltonian
but here the roles of spin and the magnetic field are played by the 
valley operator $\vec \efftau$ and the disorder-induced effective magnetic
field $\vec b$, respectively. 

In the presence of time-reversal ($T$) symmetry 
$b_z = 0$, 
because in this case $T \psi_v$ is $\psi_{-v}$ up to a phase factor, 
implying 
$|\Psi_\sigma^{(v)} (\vec r_{l\sigma})|^2 = |\Psi_\sigma^{(-v)} (\vec r_{l\sigma})|^2$
and therefore $F_{l\sigma}^{(z)} = 0$.
In CNTs, in the case of moderate magnetic fields this statement still holds:
an axial magnetic field induces an Aharonov-Bohm phase which does not 
modify the electronic wave functions (although induces energy shifts), and 
a perpendicular magnetic field interacts primarily with the spin of electrons via the
Zeeman effect as long as the magnetic length is much larger than the
nanotube radius.
Therefore, throughout this paper we omit the $b_z$ term from
the effective disorder Hamiltonian $H_{\rm dis,eff}$.

The remaining valley-Zeeman field $\vec b = (b_x,b_y,0)$ is 
random in the sense that it depends on the 
actual disorder configuration.
We consider some statistical properties of the valley-Zeeman field in the following.
First, the averages of its components are
\bnen
\langle b_{x,y} \rangle = \Omega_{\rm cell} 
\sum_{l\sigma} \langle V_{l\sigma} \rangle  F_{l\sigma}^{(x,y)}=
0.
\eden
Second, the variance of the components can be evaluated assuming
that the envelope functions are ``flat'', i.e., 
$|\Psi_\sigma^{(v)}|^2 = 1/(\Omega_{\rm cell} N)$, with
$N$ being the total number of carbon atoms in the QD.
The result is
$\langle b_{x,y}^2 \rangle = \frac{\langle V_i^2 \rangle}{2N}$.
Furthermore, it can be proven that for large $N$, the quantities $b_0$, 
$b_x$, and $b_y$ become statistically independent and 
characterized by Gaussian distributions, potentially facilitating future modeling
of carbon-based QDs where disorder-averaging is necessary.

Our results imply that the disorder-induced valley splittings in the quantum-dot
energy spectrum should have an order of magnitude 
$\sqrt{\langle b_{x,y}^2 \rangle}$.
The typical on-site energy $V_i$ on an impurity site is presumably on the atomic 
energy scale, therefore we take 1 eV as an estimate. 
Taking a quantum dot containing $10^5$ atomic sites and 50 impurity sites
we find the disorder-induced valley mixing energy scale
$\sqrt{\langle b_{x,y}^2 \rangle} \approx 50 \microev$,
consistent with recent experiments carried out on CNT single and double 
quantum dots\cite{Kuemmeth-spinorbit-in-cnts,Churchill-13cntprl}.

It is important to note that the model presented in this section relies on the 
assumption that the disorder-induced valley-mixing energy scale is much smaller
than the level spacing in the QD, i.e., the energy distance between the 
fourfold-degenerate level under consideration and its neighboring fourfold-degenerate
levels.
Recent measurements\cite{Kuemmeth-spinorbit-in-cnts,Churchill-13cntprl} imply
that this assumption is reasonable in clean CNT QD devices.
In the case when this assumption is invalid, i.e., if the valley-mixing energy scale
becomes comparable to the level spacing, then valley mixing might become efficient
\emph{between} subsequent fourfold-degenerate levels, which implies that
even the picture of independent fourfold-degenerate levels breaks down, let
alone our model based on the concept of the valley-Zeeman field acting on a
single fourfold-degenerate level.
The mechanism that strong disorder mixes subsequent levels might actually be a 
reason for observing twofold (as opposed to fourfold) electron shell filling patterns in
a number of CNT QD experiments. 

For sake of completeness, we give the disorder-independent part of the 
single-electron Hamiltonian corresponding to a fourfold-degenerate
QD energy level.
The finite curvature of the CNT enhances spin-orbit interaction and
leads to a significant splitting ($\Dso\sim 100 \mu$eV) 
of the four levels\cite{Ando-spinorbit,Kuemmeth-spinorbit-in-cnts}.
Furthermore, an external magnetic field induces a Zeeman splitting
of the spin states, and its axial component induces a splitting of the 
valley states as well.
Therefore, the disorder-independent part of the Hamiltonian is
\bnen
\label{eq:disorderindependent}
H_0 = - \frac \Dso 2 \efftauz s_z+
\mu_B \vec B \cdot \left(
  \frac{1}{2} g_s \vec s +
  \frac{1}{2} g_v \efftauz \vec{\hat z}
\right),
\eden
where $\Dso$ describes the energy splitting caused by the curvature-enhanced
spin-orbit interaction,
$g_s$ and $g_v$ are the spin and valley  $g$ factors, respectively,
and $\vec{\hat z}$ is the unit vector in the $z$ direction.
The form of the Hamiltonian $H_0$ reflects the fact that at zero magnetic field, 
the Kramers theorem implies that the state pairs connected
by time reversal, i.e., 
($\psi_{K\uparrow},\psi_{K' \downarrow}$) 
and  
($\psi_{K\downarrow},\psi_{K' \uparrow}$) are degenerate.
In Fig. \ref{fig:singledot}(b), we give an example for the evolution of a 
fourfold-degenerate level with magnetic field, which we obtain by diagonalizing
$H_0+H_{\rm dis,eff}$ [cf. Eqs. \eqref{eq:srd} and \eqref{eq:disorderindependent}].
The parameters used for Fig. \ref{fig:singledot}(b) are\cite{Kuemmeth-spinorbit-in-cnts} 
$\Dso = 370 \microev$, $\sqrt{b_x^2+b_y^2} = 30 \microev$, $g_s =2$, and
$g_v = 54$. 
The valley-independent term proportional to $b_0$, which would shift the four levels
simultaneously, is neglected.

To conclude this section: we have demonstrated that short-range disorder
in CNT QDs appears as a random (in magnitude and direction)
 valley-Zeeman field in the effective
Hamiltonian describing a fourfold- (spin and valley) degenerate quantum-dot 
level.
We note that our derivation is not by any means specific to the particular geometry
of nanotubes, and we expect the same qualitative consequences of short-range
disorder in the case of 
electrostatically defined QDs in graphene\cite{Recher-grapheneqds}
or silicon\cite{Shaji-siliconspinblockade,Culcer-prb-silicondqds,Culcer-multivalley,
Friesen-siliconvalleyorbit}.

\section{Transport model for a double quantum dot}
\label{sec:transportmodel}

Our aim in this section is to provide a model for electronic transport through
a few-electron CNT DQD which takes into account the following characteristic features
of CNTs: 
(i) fourfold (spin and valley) degeneracy of the spectrum
(ii) spin-orbit interaction, and 
(iii) disorder-induced valley mixing.
In the subsequent sections, we use this model to calculate the leakage current
through a CNT DQD in the spin-valley blockade regime.

\subsection{Hamiltonian}

\newcommand{\hdqd}{H_{\rm DQD}}
\newcommand{\cre}{d^\dag}
\newcommand{\ann}{d}
We use a constant-interaction Hamiltonian to model the few-electron
CNT DQD.
We take into account a single fourfold (spin and valley) energy
level in each QD.
We consider the case of spin- and valley-conserving interdot tunneling.
We write the Hamiltonian in terms of 
creation $\cre_{Lvs}$ ($\cre_{Rvs}$) and 
annihilation $\ann_{Lvs}$ ($\ann_{Rvs}$) operators of
electrons on the left (right) dot having valley and spin quantum numbers
$v$ and $s$, respectively,
\begin{subequations}
 \label{eq:hdqd}
 \begin{eqnarray}
  \hdqd &=&
   H_{\rm pot} + H_{\rm e-e}+H_{\rm so} + H_{\rm dis}+H_{\rm magn} + H_{\rm tun},
   \nonumber \\
   \\
  H_{\rm pot} &=& \sum_{d=L,R} \epsilon_d n_d, \label{eq:hpot} \\
  H_{\rm e-e} &=& \frac U 2 \sum_{d=L,R}  n_d (n_d-1) + U' n_L n_R,\\
  H_{\rm so} &=& -\frac{\Dso}{2}\sum_{d=L,R} s_{d,z} \tau_{d,z},
   \label{eq:hso} \\
  H_{\rm dis} &=& \sum_{d=L,R} \left(b_{d,x} \tau_{d,x} + b_{d,y} \tau_{d,y}\right),\\
  H_{\rm magn} &=&  \mu_B \vec B \cdot \sum_{d=L,R} 
   \left(\frac{1}{2} g_s \vec s_d + 
   \frac{1}{2} g_v  \tau_{d,z} \vec{\hat z}\right), \label{eq:hmagn}\\
  H_{\rm tun} &=& t \sum_{vs} \cre_{Lvs} \ann_{Rvs} + {\rm H.c.} \label{eq:htun}
 \end{eqnarray}
\end{subequations}
The terms \eqref{eq:hpot}--\eqref{eq:htun} in the Hamiltonian describe the effects of 
electrostatic potential difference between the dots, 
electron-electron interaction,
spin-orbit coupling,
short-range disorder,
external magnetic field,
and interdot tunneling, respectively.
Here $d=L,R$ is the QD index,
$n_{dvs}=\cre_{dvs} \ann_{dvs}$, $n_d=\sum_{vs} n_{dvs}$,
$\tau_{d,i} = \sum_{v,v',s} \tau_{i,vv'} \cre_{dvs} \ann_{dv's}$,
$s_{d,i} = \sum_{v,s,s'} s_{i,ss'} \cre_{dvs} \ann_{dvs'}$,
and both $\tau_i$ and $s_i$ ($i=x,y,z$) are the three Pauli matrices.
Note that we have incorporated the valley-independent disorder-induced terms 
$\sim \tau_{L,0}$ and $\sim \tau_{R,0}$ into $\epsilon_L$ and $\epsilon_R$, 
respectively.
The operators $\tau_{d,k}$ ($d=L,R$, $k=0,x,y,z$) defined above 
are the many-body generalizations
of the single-particle operator $\tilde\tau_k$ defined in Eq. \eqref{eq:efftau} 
but in Eq. \eqref{eq:hdqd} and henceforth we suppress the tilde for simplicity.
We emphasize that the disorder-induced valley-Zeeman-fields 
$\vec b_L = (b_{L,x},b_{L,y},0)$ and
$\vec b_R = (b_{R,x},b_{R,y},0)$ are different in general, since the electrons on the
left and right dot interact with a different set of impurities and therefore feel different
disorder configurations.
This feature is reminiscent of hyperfine interaction in conventional semiconductor
DQDs, and will play a critical role in all the results we present in the forthcoming 
sections.

The system is in the spin-valley blockade regime if the available charge
configurations for transport are the (1,1), (0,2), and (0,1) 
configurations
(in general, these numbers might refer to the occupations in addition to 
completely filled shells, see below).
In this situation, the only relevant parameter of $H_{\rm pot}+H_{\rm e-e}$ is
the energy difference (or ``detuning'') $\Delta$ of (1,1) and (0,2) states: 
$\Delta = \epsilon_L - \epsilon_R + U' - U $.
All results presented in this work correspond to zero detuning, $\Delta=0$,
implying that the actual values of $\epsilon_{L}$, $\epsilon_{R}$, $U$, and $U'$ 
do not affect our results.
However, since in the following we neglect hybridization with (1,0) and (2,0), 
we implicitly assume that
 $t/(U-U') = t/(\epsilon_L-\epsilon_R) \ll 1$.
 
Motivated by the experiments we try to model here, 
we consider the case of an axial magnetic field: 
$\vec B = (0,0,B)$.
In the CNT QD studied by Kuemmeth \emph{et al.}
\cite{Kuemmeth-spinorbit-in-cnts} a spin $g$ factor $g_s \approx 2$
and a valley $g$ factor $g_v \sim 50$ have been found.
However, the results we present in this work are insensitive to these values, because
(at least in the parameter regimes under consideration here)
(i) spin-Zeeman splitting do not affect the dynamics and
(ii) we plot the magnetotransport curves against the field-induced valley splitting
$\Delta_v = g_v \mu_B B$ and not against the field $B$ itself.

Our constant-interaction approximation has the advantage of simplicity
but also has the drawback that it does not account
for the recently predicted Wigner-molecule formation effect
\cite{Roy-coulombincnt,Wunsch-coulombincnt,Secchi-coulombincnt}.
This restricts the applicability of our model to 
(i) short quantum dots, where the confinement energy exceeds the interaction
energy, or 
(ii) DQD systems where the environment (the metallic gate electrodes or
the dielectric substrate, 
for example) provides a strong electrostatic screening and hence weakens the 
electron-electron interaction.
Wigner-molecule formation implies a strong suppression of the 
supersinglet-supertriplet gap in the (0,2) charge configuration, which
suppresses the Pauli blockade as well.
The fact that Pauli blockade has been observed
\cite{Buitelaar-spinblockade,Churchill-cntspinblockade,Churchill-13cntprl,
Chorley-cntspinblockade} in CNT DQDs indicates that
the samples used in those experiments are closer to the constant-interaction 
regime than to the Wigner-molecule regime, which is a further motivation for us
to use the constant-interaction model in our calculations.

Our Hamiltonian does not contain hyperfine interaction
and spin- or valley-flip interdot tunneling matrix elements, although 
hyperfine interaction\cite{Jouravlev-spinblockade,Koppens-spinblockade} and 
spin-orbit-induced
spin-flip tunneling\cite{Danon-spinblockade,NadjPerge-spinblockade} 
have proven to be important in the understanding of 
spin blockade experiments in conventional semiconductor quantum dots.
We neglect hyperfine interaction in this work because 
theoretical estimates indicate that its characteristic energy scale is below $5$neV
even for fully $^{13}$C-enriched 
samples\cite{Yazyev-graphenehf,Fischer-prb-cnt,Palyi-spinblockade}),
being small compared to other relevant energy scales in our system
i.e., spin-orbit splitting $\gtrsim 100\mu$eV, disorder $\gtrsim 10\mu$eV, interdot
tunneling, and valley splitting $\gtrsim \mu$eV, see forthcoming sections.
Note that recent experiments\cite{Churchill-cntspinblockade,Churchill-13cntprl} 
indicate a two orders of magnitude larger hyperfine energy scale than the theoretical
estimates, and therefore
we cannot be conclusive about the relevance of this effect.
Experiments in GaAs and CNT DQDs have shown that hyperfine coupling becomes 
especially relevant at suppressed interdot tunneling or large detuning, therefore our
model excluding this mechanism might not be adequate in that regime.

Spin-orbit-induced spin-flip interdot tunneling could in principle be present in our
system, but only between states having a $\pm 1$ difference in their circumferential 
quantum number, as it can be deduced from Eqs. (31)-(33) of 
Ref. \onlinecite{Bulaev-socincntdots}. 
This possibility is not ruled out in some of the spin-valley blockade 
transport experiments\cite{Churchill-cntspinblockade} as those were
not performed in the actual (1,1)-(0,2)-(0,1) regime (which would imply that
in both dots the electrons occupy the lowest-energy circumferential mode of the CNT
and therefore spin-flip tunneling is forbidden).
For example, in the $(4n+1,4m+1)$-$(4n,4m+2)$-$(4n,4m+1)$ regime
the spin-valley blockade could take place ``on top of'' $n$ ($m$) filled shells
in the left (right) dot.
However, we postpone the analysis of spin-flip tunneling for future work and in
Sec.~\ref{sec:strongsoi} we demonstrate that agreement with experimental 
results can be obtained from our model even though spin-flip tunneling is not
taken into account.

\subsection{Generalized master equation}

We apply the master-equation formalism to describe the transport process
through the serially coupled DQD system.
The DQD charge configurations which are relevant for the transport process
considered here are the (1,1), (0,2), and (0,1) configurations.
Hence the state of the DQD system is described by the $26\times 26$ density 
matrix $\rho$, where the Hilbert space is spanned by
16 states in the (1,1) charge configuration, six states in the (0,2)
charge configuration, and four states in the (0,1) configuration.
The time dependence of  $\rho$ is governed by the 
generalized master equation or Lindblad equation
\bnen
\dot\rho=-\frac i \hbar [\hdqd,\rho]+D\rho.
\eden

The dissipative term $D\rho$ describes the tunneling events to and from the DQD, 
characterized by the rates 
\newcommand{\ratel}{\Gamma_L}
\newcommand{\rater}{\Gamma_R}
$\ratel$ and $\rater$, respectively.
It has the following form:
\bean \nonumber
D\rho &=& 
\ratel \sum_{vs}  
\left(
  \cre_{Lvs} \rho \ann_{Lvs}
  - \frac 1 2 \rho \ann_{Lvs} \cre_{Lvs} 
  - \frac 1 2 \ann_{Lvs} \cre_{Lvs} \rho 
\right)\\
&+&
\rater \sum_{vs} 
\left(
  \ann_{Rvs} \rho \cre_{Rvs}
  - \frac 1 2 \rho \cre_{Rvs} \ann_{Rvs} 
  - \frac 1 2 \cre_{Rvs} \ann_{Rvs} \rho
\right) \nonumber. \\
\eean
Here the creation and annihilation operators are restricted to the charge 
configurations participating in the transport process.

\subsection{Secular approximation}

We assume that the splittings between the eigenvalues of $\hdqd$ are larger
than the level broadenings set by the tunneling energies
$h \Gamma_L$ and $h \Gamma_R$.
This allows us to use the so-called secular 
approximation\cite{Breuer-openquantumsystems},
i.e., to assume that the steady-state density matrix is diagonal in the
eigenbasis of $\hdqd$.
Hence the generalized master equation
 simplifies to a steady-state classical master equation (CME),
\begin{subequations}
 \label{eqs:cme}
 \begin{eqnarray}
  0 &=& \dot\rho_\alpha =
   -\rho_\alpha \rater \sum_j p_{j\alpha} 
   +\ratel \sum_j \rho_j r_{\alpha j}, \label{eq:cme2} \\
  0 &=& \dot\rho_i =
   -\rho_i \ratel \sum_\beta r_{\beta i}
   + \rater \sum_\beta \rho_\beta p_{i\beta} \label{eq:cme1}.
 \end{eqnarray}
\end{subequations}
Here $\alpha, \beta = 1,\dots,22$ 
($i,j=1,2,3,4$) refers to the two-electron (single-electron)
DQD energy eigenstates, $\rho_\alpha = \rho_{\alpha \alpha}$ and
$\rho_i = \rho_{ii}$,  and
\begin{subequations}
  \bean
    r_{\alpha i} &=& \sum_{vs} \left| \langle i |
      \ann_{Lvs}
      | \alpha \rangle
      \right|^2,\\
    p_{i \alpha} &=& \sum_{vs} \left| \langle i |
      \ann_{Rvs}
      | \alpha \rangle
      \right|^2. \label{eq:decayrates}
  \eean
\end{subequations}  

\subsection{Eliminating (0,1) states from the classical master equation}

The Hamiltonian has a block-diagonal structure: the two-electron 
[(1,1) and (0,2)] and single-electron (0,1) blocks are uncoupled.
However, the Lindblad terms do couple these sectors because they describe
single-electron tunneling onto and from the DQD.
The coupling is appearing in the CME in the form
of  the rates
$\ratel r_{\alpha i}$ 
and 
$\rater p_{i \alpha}$.

Throughout this analysis, we consider the case $\ratel \gg \rater$.
The reason is that in the spin-valley blockade regime the characteristic scale of the 
rates  $\ratel r_{\alpha i}$ are largely independent of the
spin and valley physics inside the DQD, whereas the rates $\rater p_{i \alpha}$ are
sensitive to those,
so in order to have the transport via the DQD sensitive to spin and valley effects,
the outgoing rates $\rater p_{i \alpha}$ should provide the transport bottleneck.

\newcommand{\redrho}{\rho}
We claim that in this limit $\rater/\ratel \to 0$, the steady-state
CME is reduced to 
a homogeneous linear set of equations $M\redrho = 0$ for the vector 
$\redrho=(\redrho_1,\redrho_2,\dots,\redrho_{22})$ which
contains the  diagonal elements
of the two-electron sector  of the DQD density matrix $\rho$, 
and the normalization condition $\sum_{\alpha} \redrho_\alpha = 1$.
The coefficient matrix $M$ is given as
\bnen
\label{eq:2ecme}
M_{\alpha\beta} = 
\sum_j 
\left(
\frac{ r_{\alpha j} p_{j \beta} }{ \sum_\gamma r_{\gamma j} }
-\delta_{\alpha \beta} p_{j\alpha}
\right).
\eden
The proof of this statement is a straightforward calculation starting from
the steady-state CME in Eq. \eqref{eqs:cme}.

Having the steady-state occupation probabilities $\redrho_\alpha$ 
and the corresponding energy eigenstates $|\alpha \rangle$ at hand,
we calculate the current as the average decay rate of the two-electron states with
respect to the steady-state distribution,
\bnen
\label{eq:current}
I = e \rater\sum_\alpha \redrho_\alpha \sum_j p_{j\alpha}.
\eden


\section{Strong spin-orbit coupling}
\label{sec:strongsoi}
\newcommand{\tsplus}{T_+^{s}}
\newcommand{\tsminus}{T_-^{s}}
In this section, we describe the spin-valley blockade effect in a 
CNT DQD in the case when 
spin-orbit coupling dominates the energy spectrum over disorder, interdot tunneling,
and magnetic-field-induced spin and valley splitting, i.e., 
$\Dso \gg b,t,\Delta_v,\Delta_s$.
Here $b$ denotes the typical energy scale of the disorder-induced
valley-Zeeman fields on the two dots.
The main result of this section is that we identify a parameter regime
($t \lesssim \frac{b_-^2}{\Dso}$, where $b_-^2 = b_L^2-b_R^2$)
where the current as the function of magnetic field (the ``magnetotransport curve'')
shows a dip around zero field, and the width of the dip is controllable
by the interdot tunneling amplitude $t$.
This field-induced increase in the current is in qualitative agreement with experiments
\cite{Churchill-cntspinblockade,Churchill-13cntprl}.
We interpret this result using L\"owdin perturbation theory\cite{Lowdin-partitioning},
and provide an analytical formula for the current which can be well
fitted to the numerical results using a single fitting parameter,
the average number of transmitted electrons between two blocking events
\cite{Qassemi-spinblockade}.
In the following, we describe the case
$t \sim \frac{b_-^2}{\Dso}$.
In Appendix \ref{app:strongdisorder}, we argue that the findings of this regime
can be extended to the regime $t \ll \frac{b_-^2}{\Dso}$ as well,
and in Appendix \ref{app:strongdisorder2} we show that
 they do not hold if $t \gg \frac{b_-^2}{\Dso}$.

\begin{figure}
\includegraphics[scale=0.23]{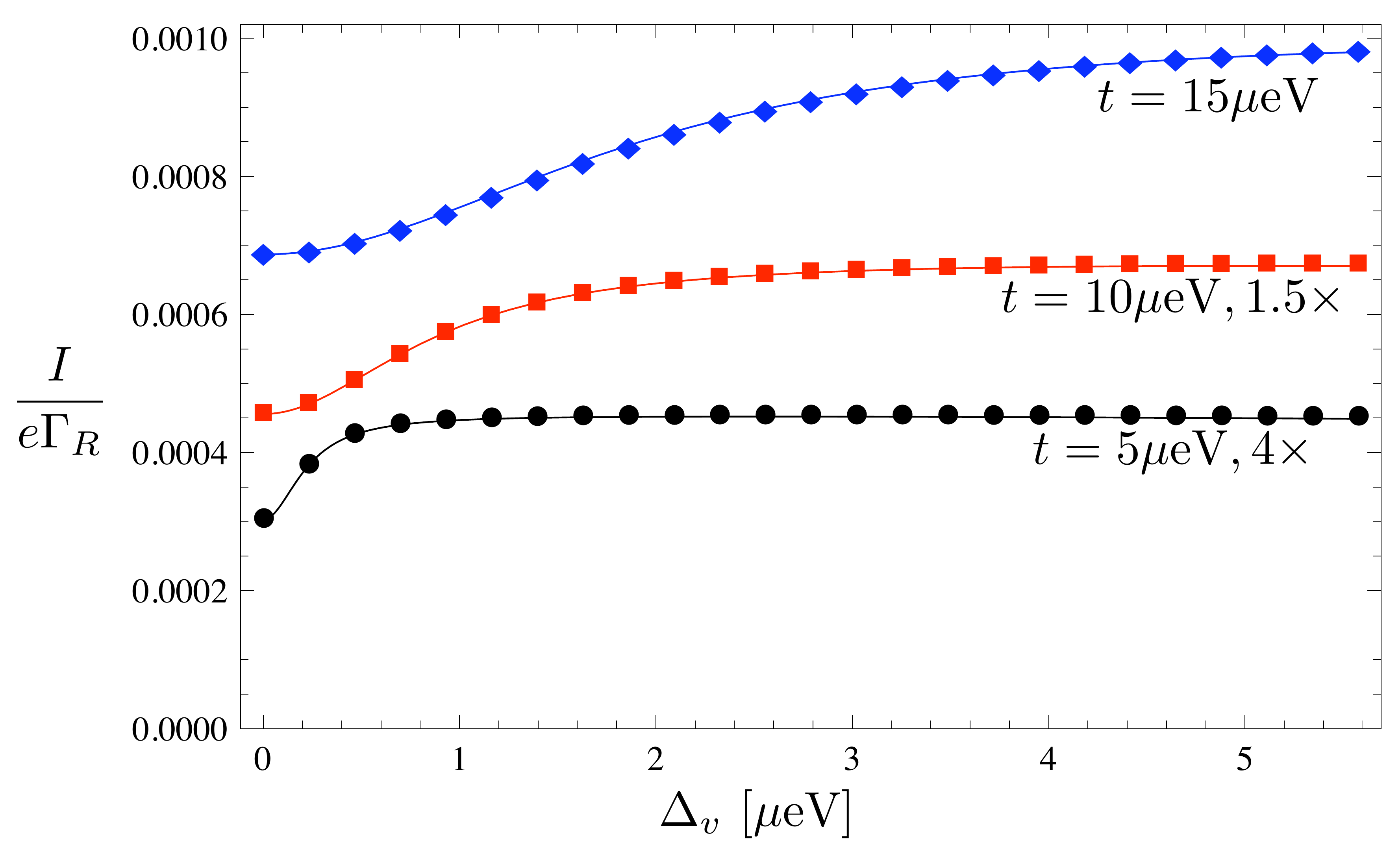}
\caption{\label{fig:dip}
(Color online)
Numerical results for the current as a function of magnetic-field-induced valley splitting
for different values of interdot tunneling (shown).
Further parameters:
$\Dso = 250 \microev$,
$b_{Lx}= 20 \microev$,
$b_{Ly} = 10 \microev$,
$b_{Rx} = 80 \microev$,
$b_{Ry} = 0 \microev$,
and therefore $b_-^2/\Dso = 23.6$.
Points: numerical data. Curves: the analytical formula \eqref{eq:burstcurrent}
fitted to the numerical data with $n^*$ as the single fitting parameter.
Lower two data sets are scaled as shown.}
\end{figure}

We start our analysis by presenting the numerical results for this regime.
In Fig. \ref{fig:dip}, we show the current as a function of the magnetic-field-induced
valley splitting $\Delta_v$, for a fixed value
of spin-orbit coupling $\Dso$ and disorder-induced valley fields 
$\vec b_L$ and $\vec b_R$ (see caption)
but different values of interdot tunneling $t$.
All parameters have a realistic order of 
magnitude\cite{Kuemmeth-spinorbit-in-cnts,Churchill-13cntprl}.
(Note that the Zeeman spin splitting $\Delta_s$ plays no role in the
transport process, see below.)
In qualitative agreement with recent 
experiments\cite{Churchill-cntspinblockade,Churchill-13cntprl}, the
data in Fig. \ref{fig:dip} shows a zero-field dip in the current, and the width of the
dip is controlled by the interdot tunneling $t$.
In all the three cases displayed, the ratio of the zero-field current 
$I_0 \equiv I(B=0)$ and the maximal current $I_{\rm max}$ 
is $I_{\rm max}/I_0 \approx 1.5$. This ratio agrees well with that observed 
experimentally in Ref.~ \onlinecite{Churchill-13cntprl} [see Fig. 3(a) therein], 
however, in Ref.~ \onlinecite{Churchill-cntspinblockade} a ratio of 
$I_{\rm max}/I_0\sim 50$ has been found [see Fig. 3(e) therein].
Below we argue that the factor $I_{\rm max}/I_0\approx 1.5$ we deduce from
Fig. \ref{fig:dip} is a rough upper bound for this quantity in the parameter regime
under consideration, and therefore we conclude that our results
(i) agree very well with the measurement of Ref.~ 
\onlinecite{Churchill-13cntprl}, and
(ii) match the measurement of 
Ref.~  \onlinecite{Churchill-cntspinblockade}
only qualitatively, which might be due to mechanisms missing from our model or 
sample parameters in the experiment
not fitting into the parameter regime we consider here. 
Further discussion on this discrepancy with 
Ref.~ \onlinecite{Churchill-cntspinblockade} 
is provided in Sec. \ref{sec:discussion}.
In the remaining part of this section, we provide an interpretation of the 
numerical results shown in Fig. \ref{fig:dip}
and derive an analytical formula for the current using
L\"owdin perturbation theory.

\begin{table*}
\begin{ruledtabular}
\begin{tabular}{cccc}
  Spin-orbit energy &
    Up-spin ($S_z=+1$)&
    Down-spin ($S_z=-1$)&
    Mixed spin ($S_z=0$)\\
  \hline
  $\Dso$ &
    $ | \qu, \qu \rangle$ & 
    $ | \kd, \kd \rangle$ & 
    \begin{tabular}{c}
        $\frac{1}{\sqrt{2}} \left(| \kd, \qu \rangle \pm | \qu, \kd \rangle \right)$ \\
        $| 0,\kd \qu \rangle $
    \end{tabular} \\
  \hline
  $0$ & 
    \begin{tabular}{c}
        $\frac{1}{\sqrt{2}} \left(| \ku, \qu \rangle \pm | \qu, \ku \rangle \right)$ \\
        $| 0,\ku \qu \rangle $
    \end{tabular} & 
    \begin{tabular}{c}
        $\frac{1}{\sqrt{2}} \left(| \kd, \qd \rangle \pm | \qd, \kd \rangle \right)$ \\
        $| 0,\kd \qd \rangle $
    \end{tabular} & 
    \begin{tabular}{c}
        $\frac{1}{\sqrt{2}} \left(| \ku, \kd \rangle \pm | \kd, \ku \rangle \right)$ \\
        $| 0,\ku \kd \rangle $ \\
        $\frac{1}{\sqrt{2}} \left(| \qu, \qd \rangle \pm | \qd, \qu \rangle \right)$ \\
        $| 0,\qu \qd \rangle $
    \end{tabular} \\
  \hline
  $-\Dso$ &
    $ | \ku, \ku \rangle$  &
    $ | \qd, \qd \rangle$  &
    \begin{tabular}{c}
        $\frac{1}{\sqrt{2}} \left(| \ku, \qd \rangle \pm | \qd, \ku \rangle \right)$ \\
        $| 0,\ku \qd \rangle $
    \end{tabular} \\
\end{tabular}
\end{ruledtabular}
\caption{\label{tab:class}
The 22 basis states used for perturbation calculations in the presence of strong
spin-orbit coupling.
The six (1,1) states involving a minus sign are supersinglets, the ten further 
(1,1) states are supertriplets. 
The six (0,2) states are supersinglets.}
\end{table*}

The transition rates in the classical master equation [Eq. \eqref{eqs:cme}]
are determined by the eigenstates of the two-electron Hamiltonian.
To provide an interpretation of the numerical results, we will describe those energy
eigenstates using perturbation theory.
We start with the two-electron Fock basis based on the single-particle states
$\psi_{\ku}$, $\psi_{\qd}$ and $\psi_{\kd}$, $\psi_{\qu}$ [the pairs are energetically 
separated by the spin-orbit energy $\Dso$ at zero field, 
see Eq. \eqref{eq:disorderindependent}].
The (1,1) states are denoted in the form $|\ku,\qu\rangle$ whereas the 
(0,2) states in the form $|0,\ku \qu \rangle$.
We perform a basis transformation in order to obtain basis states which are
eigenstates of the two-electron spin-orbit Hamiltonian [Eq. \eqref{eq:hso}] 
and have well-defined  supersinglet or supertriplet character at the same time. 
This new basis is presented in Table \ref{tab:class}, classified according to their
properties outlined below.
This basis will serve as the set of unperturbed states in our perturbation calculations.

An important simplifying observation is that even in the presence
of spin-orbit coupling and a magnetic field parallel to the nanotube axis,
the axial component of the electron spin $S_z$ is conserved.
This allows us to separate the 22 states of the two-electron basis
to three uncoupled spin subspaces (see columns in Table \ref{tab:class}): 
five states which are spin polarized with a polarization aligned with the $z$ axis
(up-spin states), 
five states which are spin polarized with a polarization antialigned with the $z$ axis
(down-spin states),
and 12 states having mixed spin states.
As the three different spin subspaces shown in the columns of Table \ref{tab:class}
are not coupled by any terms in the Hamiltonian, 
the Zeeman spin splitting $\Delta_s$ plays no role in the transport process.
Besides their spin state, our unperturbed states can also be classified according to 
their spin-orbit energy.
Five (five) of those have a spin-orbit energy $\Dso$ ($-\Dso$), and
12 have a vanishing spin-orbit energy (see rows in Table \ref{tab:class}).

\begin{figure*}
\includegraphics[scale=0.5]{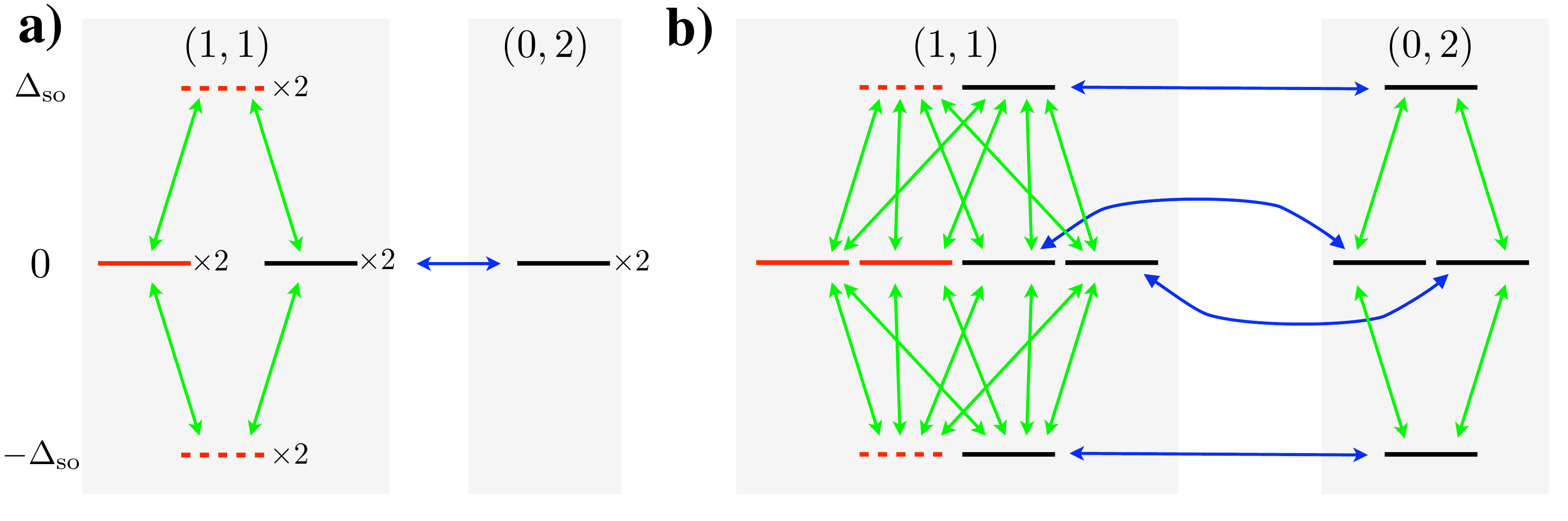}
\caption{\label{fig:leveldiagram}
(Color online)
Unperturbed two-electron states (lines) and their energies in a CNT DQD at strong 
spin-orbit coupling and zero magnetic field  (cf. Table \ref{tab:class}).
Horizontal arrangement of the states reflects charge configuration and
vertical arrangement reflects spin-orbit energies. 
Red/gray lines: supertriplet states.
Dashed red/gray lines: blocked supertriplet states.
Black lines: supersinglet states. 
Green/light gray (blue/dark gray) arrows correspond to off-diagonal elements of the Hamiltonian, 
induced by disorder (interdot tunneling).
(a) Up-spin and down-spin states. The indicated $\times 2$ degeneracy corresponds
to the two possible spin configurations. Different spin species are uncoupled.
(b) Mixed-spin states.
}
\end{figure*}

To visualize the matrix elements of the Hamiltonian,
in Fig. \ref{fig:leveldiagram} we show the level diagram
of the unperturbed basis states we introduced in Table \ref{tab:class}.
The horizontal arrangement of the states reflects the charge configuration, and
the vertical arrangement reflects the spin-orbit energies. 
Red/gray lines denote supertriplet states and black lines denote supersinglet
states. 
The green/light gray (blue/dark gray) arrows correspond to off-diagonal elements of the Hamiltonian
in this basis, induced by disorder (interdot tunneling).

In the polarized spin subspaces [Fig. \ref{fig:leveldiagram}(a)]
 the high- and low-energy
(1,1) states [dashed lines in Fig. \ref{fig:leveldiagram}(a)] 
are coupled to the (0,2) state via 
disorder and tunneling, resulting in a small decay rate at small fields
($\Delta_v \ll \Dso$),
\bnen
  \Gamma_{s, \pm} = 
  \Gamma_s \left(1 \mp 2\frac{\Delta_v}{\Dso}\right)^{-4},
\eden
where
the $\pm$ sign refers to the up-spin and down-spin subspaces, respectively,
 $\Gamma_s = \frac{t^2 b_a^2}{\Dso^4} \Gamma_R \ll \Gamma_R$ and
$\vec b_a = \vec b_L - \vec b_R$.
Here and hereafter the decay rate of a two-electron state $\alpha$
is meant to be the sum of the four transition rates into the four different (0,1)
single-electron states, i.e., $\sum_{j=1}^4 p_{j\alpha}$.
In the up-spin (down-spin) subspace the decay rate increases (decreases)
as the magnetic field increases because the magnetic field
pushes the high- and low-energy states closer to (away from) the
zero-energy (1,1) and (0,2) states, cf. Eq. \eqref{eq:hmagn}.
The decay rate $\Gamma_s$ is fourth order in small parameters, therefore
we call these four states 
(i.e., the high- and low-energy up-spin and down-spin states) ``blocked''.

To describe the energy eigenstates in the spin-polarized zero-energy subspace,
conventional degenerate perturbation theory is not applicable since
the perturbative hybridization of the (1,1) supertriplet state with the (1,1) supersinglet
would include a zero-energy denominator.
Therefore, we apply L\"owdin perturbation theory\cite{lowdinfootnote}
to derive an effective Hamiltonian for the zero-energy spin-polarized subspace.
At zero field, we obtain
\bnen \label{eq:h0pm}
H_{0,\pm}=
\left(\bna{ccc}
  0   &
    \mp \frac{b^2_-}{\Dso} &  
    0 \\
  \mp \frac{b^2_-}{\Dso} &  
    0 &
    \sqrt{2}  t \\
  0 &
    \sqrt{2} t &
    0
\eda\right).
\eden
In $H_{0,\pm}$, the first index refers to the zero-energy subspace and $\pm$ to
the up-spin and down-spin subspaces. 
The effective Hamiltonian $H_{0,\pm}$ corresponds to the following ordering of the
basis states:
(1,1) supertriplet, (1,1) supersinglet, (0,2) supersinglet.
Remarkably, $H_{0,\pm}$ is independent of the
angle between the two disorder-induced valley fields in the double dot.
From Eq. \eqref{eq:h0pm} and our assumption $\frac{b_-^2}{\Dso} \sim t$,
it follows that
the three basis states are completely mixed, and each of them acquires a
decay rate $\sim \Gamma_R$.
Therefore with respect to the spin-polarized subspaces, we conclude that in the regime
considered in this section, the ten energy eigenstates can be divided to 
a set of four blocked states decaying with slow rates 
$\Gamma_{s,\pm} \ll \Gamma_R$,
and six unblocked states which decay orders-of-magnitude faster 
(with rates $\sim \Gamma_R$) than the blocked ones.

Now we extend this analysis to the 12-dimensional mixed-spin subspace
[Fig. \ref{fig:leveldiagram}(b)]. 
In the high- and low-energy mixed spin subspaces, the effective Hamiltonian 
we obtain is (common diagonal elements are omitted)
\bnen \label{eq:hpm0}
H_{\pm,0}=\left(\bna{ccc}
  0 &
    -\frac{2 b^2_- \Delta_v}{\Dso^2-4\Delta_v^2}  &
    0 \\
    -\frac{2 b^2_- \Delta_v}{\Dso^2-4\Delta_v^2}  &
    0 &
    \sqrt{2} t \\
  0 &
    \sqrt{2} t  &
    \mp \frac{b^2_- \Dso}{\Dso^2-4\Delta_v^2}
\eda\right).
\eden
Here the ordering of states is analogous to that in Eq. \eqref{eq:h0pm},
and the first index of $H_{\pm,0}$ refers to the high- or low-energy subspace
whereas the second index refers to the mixed spin subspace.
At zero field, the valley splitting is $\Delta_v = 0$, implying that the first
basis state, i.e., the (1,1) supertriplet is uncoupled from the other
two states, in particular, from the (0,2) supersinglet.
Therefore, at zero field the high- or low-energy (1,1) supertriplet state 
[dashed lines in Fig. \ref{fig:leveldiagram}(b)] can decay
only due to its perturbative coupling (via disorder and tunneling) to the 
two zero-energy (0,2) supersinglets.
From L\"owdin theory, we infer that the decay rate of the high- and low-energy
(1,1) supertriplets due to these processes is
\bnen
\Gamma_{s,0} = \frac 1 2 \left(\Gamma_{s,+} + \Gamma_{s,-}\right),
\eden
and therefore these two states are blocked in the sense defined above.
However, according to Eq. \eqref{eq:hpm0}, a finite valley splitting $\Delta_v$ 
induces mixing between the two (1,1) states.
This mixing provides an additional decay channel for the (1,1) supertriplet state
with a rate 
\bnen
\label{eq:crossoverrate}
 \Gamma_{c} = \frac{2 b_-^4 \Delta_v^2}
  {t^2 \left(\Dso^2 - 4 \Delta_v^2 \right)^2}\Gamma_R,
\eden
inferred using standard perturbation theory in the field-induced coupling term.
This rate becomes much faster than the slow rate $\Gamma_{s,0}$ if the field
is strong enough to ensure $\Delta_v > \frac{t^2 b_a}{\sqrt 2 b_-^2}$.
In conclusion, we have found that in both the high- and low-energy mixed spin
subspace the total decay rate of the (1,1) supertriplet state 
$\Gamma_{s,0}+\Gamma_c$ changes dramatically as
the magnetic field is turned on: at zero field these two states are blocked, having the
slow decay rate $\Gamma_{s,0}$ whereas at finite field their decay rate grows with
orders of magnitudes. 

In the six-dimensional zero-energy mixed spin subspace, the order of magnitude
of the decay rates is not influenced by the magnetic field.
Each of these states decay fast compared to the slow rate $\Gamma_s$.
This can be derived in the same way as shown at the discussion of the zero-energy 
spin-polarized subspace and Eq. \eqref{eq:h0pm}.

Using the explicitly calculated decay rates, we can set up a semiphenomenological
analytical formula for the current.
To this end, we regard the transport process as an alternation of charge transfer 
``bursts'' (subsequent occupation of unblocked states) and blocking events 
(due to occupying one of the blocked states).
We assume that a burst corresponds to a transfer of $n^*$ electrons on average, 
i.e., $n^*$ is not necessarily integer\cite{Qassemi-spinblockade}.
Since the charge bursts happen fast compared to the time spent in a blocked state,
the average time between two subsequent bursts can be estimated as
the average of the decay times of the six blocked states, i.e., 
\bnen
\label{eq:bursttime}
T_{\rm burst} \approx \frac 1 6 \left[2 \Gamma_{s,+}^{-1}
 + 2 \Gamma_{s,-}^{-1} + 2 \left(
\Gamma_{s,0} + \Gamma_c
\right)^{-1}\right].
\eden
As a burst transfers $n^*$ electrons on average, the current can be
expressed as 
\bnen
\label{eq:burstcurrent}
I \approx \frac{e n^*}{T_{\rm burst}}.
\eden
Equation \eqref{eq:burstcurrent}, together with Eq. \eqref{eq:bursttime}
and the decay rates calculated above, provides an analytical expression
for the current as a function of the parameters of the Hamiltonian and $\Gamma_R$,
having a single phenomenological parameter $n^*$.
We have fitted this analytical result using $n^*$ as the single fitting parameter 
to our numerical results (Fig. \ref{fig:dip}), and we have found 
$n^* \approx 3.2$ irrespective of the value of tunneling amplitude $t$.
As seen in Fig. \ref{fig:dip}, 
this value of $n^*$ gives an excellent agreement between our numerical
and analytical results in the considered range of magnetic field.
By repeating the numerical calculations and the fittings for various
disorder configurations we generally find good agreement between
numerics and analytics.
The values we obtain for $n^*$ are typically between $1.4$ and $5.2$,
indicating that $n^*$ is not universal but depends on the details of the Hamiltonian.

Our analytical result for the current enables us to qualitatively explain
two characteristic features of the magnetotransport curves shown in
Fig. \ref{fig:dip}.
One of those features is the ratio $I_{\rm max}/I_0 \approx 1.5$.
Evaluating the current according to Eq. \eqref{eq:burstcurrent}
at zero field, we find 
$I_0 = e n^* \Gamma_s$,
whereas at high field, where $\Gamma_c \gg \Gamma_s$,
we can neglect 
$\left(\Gamma_{s,0}+\Gamma_c\right)^{-1}$ in Eq. \eqref{eq:bursttime}
and find $I_{\rm max} \approx en^* 6 \Gamma_s/4$,
resulting in the ratio $I_{\rm max}/I_0 \approx 1.5$,
in correspondence with our numerical results in Fig. \ref{fig:dip} and the
experimental data of Ref.~ \onlinecite{Churchill-13cntprl}.
For this estimate we neglected the field dependence of the rates $\Gamma_{s,\pm}$
but taking that into account could only lower the ratio $I_{\rm max}/I_0$.
A second feature observed in Fig. \ref{fig:dip} is that the width of the zero-field dip
of the magnetotransport curve depends on the tunneling amplitude $t$.
This is explained by the fact that the crossing-over rate in
Eq. \eqref{eq:crossoverrate} depends
on the tunneling amplitude as $\Gamma_c \propto 1/t^2$,
i.e., the stronger the tunneling, the ``slower'' the crossover of $\Gamma_c$ as
the magnetic field increases, and therefore the wider the zero-field dip
in the magnetotransport data. 
In Appendix \ref{app:strongdisorder} we argue that the conclusions drawn in 
this section for the case $t \sim \frac{b_-^2}{\Dso}$ can be generalized to the
regime $t\ll \frac{b_-^2}{\Dso}$, and therefore the range of validity
of our results is actually $t\lesssim \frac{b_-^2}{\Dso}$.

Finally we point out a possible generalization of our results. 
In the system under consideration, the spin and valley degrees of freedom
play a symmetric role in
the absence of disorder and magnetic field, since the spin-orbit Hamiltonian 
$H_{\rm so} \propto s_z \tau_z$ is symmetric in spin and valley,
and the interdot tunneling conserves both spin and valley. 
The results of this section show that if disorder provides an inhomogeneous 
valley-Zeeman field (coupled to $\tau_x$ and $\tau_y$) in the DQD, then
the dynamics becomes independent of the spin-Zeeman splitting, and 
the magnetotransport curve shows a dip at zero axial magnetic field. 
These results can be transferred to the case when the role of spin and valley
are exchanged:
in the hypothetic case of absence of disorder, an inhomogeneous spin-Zeeman field,
coupled to $s_x$ and $s_y$ but not to $s_z$, e.g., coming from a 
perpendicular-to-nanotube-axis magnetic field, would imply that 
the dynamics becomes independent of the valley-Zeeman splitting, and 
the magnetotransport curve would show a dip at zero axial magnetic field.

To conclude this section: solving the transport master equation numerically,
we have found that if $\Dso \gg \frac{b_-^2}{\Dso} \gtrsim t$
and $\Dso \gg \Delta_v$, then the magnetotransport curves 
show a zero-field dip with a width
that is controllable by the interdot tunneling amplitude $t$,
in agreement with a recent experiment\cite{Churchill-cntspinblockade}.
Using L\"owdin perturbation theory, we gave an analytical formula for the current and 
based on that, a qualitative interpretation of the features of our numerical results.
We emphasize that the observed characteristic magnetotransport pattern
is due to the different disorder-induced effective valley-Zeeman fields on
the two quantum dots. 
Our theory predicts a typical ratio of the finite-field and zero-field
currents $I_{\rm max}/I_0 \lesssim 1.5$, which is in line with the
experimental result of Ref.~ \onlinecite{Churchill-13cntprl},
but different from that of Ref.~ \onlinecite{Churchill-cntspinblockade},
possibly due to experimental sample parameters not fitting into the parameter
range studied here or mechanisms missing from our transport model.


\section{Strong disorder}
\label{sec:strongdisorder}
In this section, we describe the spin-valley blockade effect in a CNT DQD in the
case of strong disorder, weak interdot tunneling and 
weak spin-orbit coupling ($b \gg t,\Dso$).
In recent experiments on clean nanotube QDs, the spin-orbit splitting of the 
fourfold-degenerate ground-state energy level has
been found significantly larger than the valley mixing energy scale. 
However, in nanotubes with stronger impurity contamination (larger radius) 
the disorder (spin-orbit interaction) energy scale is expected to increase (decrease),
and the regime considered in this section might be reached. 
A further motivation to study this regime is its possible relevance for certain 
silicon\cite{Shaji-siliconspinblockade,Culcer-prb-silicondqds,Culcer-multivalley,
Friesen-siliconvalleyorbit} or
graphene-based quantum dots\cite{Recher-grapheneqds}. 
In those material systems, the spin-orbit interaction is expected to be
smaller than in CNTs but short-range disorder couples valleys for the same reason
as it does in CNTs.

As the main result of this section, we show that in the parameter regime under
consideration, the magnetotransport curve 
shows a zero-field dip or peak depending on the disorder configuration. 
We find that at a given value of the external magnetic field, 
the current is determined by three parameters (if interdot tunneling
$t$ and emptying rate $\Gamma_R$ are fixed): 
the angle $\theta^{\rm (tot)}$ between 
the two total valley-Zeeman fields 
$\vec b_L^{\rm (tot)} \equiv \vec b_L + \Delta_v \vec{\hat{z}}/2$ and
$\vec b_R^{\rm (tot)} \equiv \vec b_R + \Delta_v \vec{\hat{z}}/2$ 
on the two dots,
and the lengths of these valley-Zeeman 
fields $b_L^{\rm (tot)}$ and $b_R^{\rm (tot)}$.
Our analysis is analogous to the derivation of the spin blockade
leakage current induced by hyperfine interaction in GaAs double 
dots\cite{Jouravlev-spinblockade}, with the most important difference being that
in our case spin-independent disorder provides a blockade-lifting mechanism via
the valley dynamics, whereas in conventional spin blockade, the
hyperfine interaction affecting spin dynamics is responsible for lifting the blockade.

\begin{figure}
\includegraphics[scale=0.23]{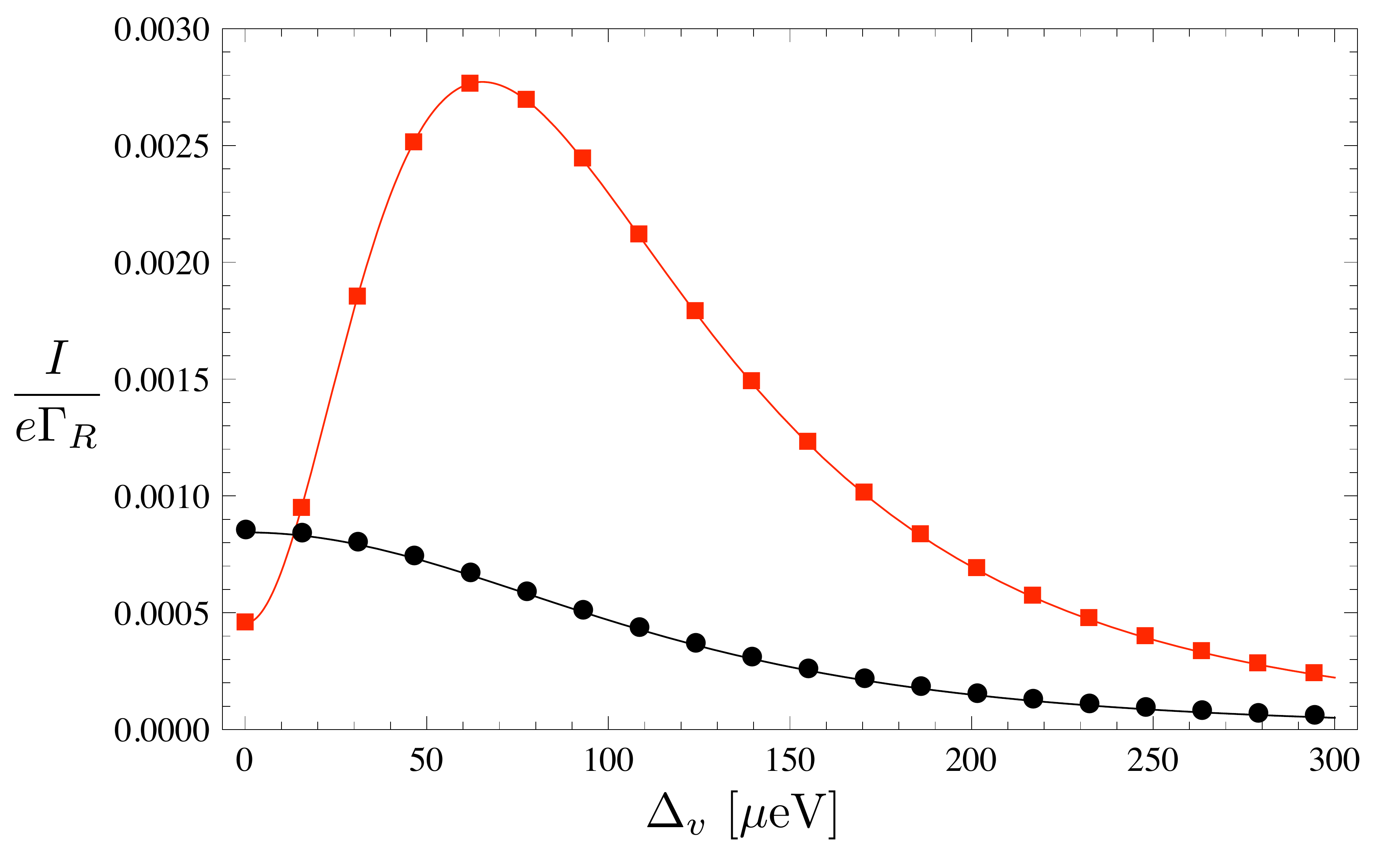}
\caption{\label{fig:strongdisorder}
(Color online)
Numerical (dots) and analytical (lines) results for the current as a function of 
magnetic-field-induced valley splitting for two different disorder realizations.
Parameters: $t=10 \microev$,
$b_{Lx} = \cos \theta\times 100 \microev$, 
$b_{Ly} = \sin \theta\times 100 \microev$,
$b_{Rx} = 200 \microev$ and
$b_{Ry} = 0 \microev$.
Circles: $\theta= \pi/4$. Boxes: $\theta= 15\pi / 16$.}
\end{figure}

We start our analysis by presenting numerical results in Fig. \ref{fig:strongdisorder}.
The figure shows the magnetic-field dependence of the current for two
different disorder realizations.
The two curves are qualitatively different: one shows a zero-field dip whereas the
other shows a zero-field 
peak.
In the following, using standard perturbation theory and an approximative
analytical solution of the master equation we show that the qualitative
difference between the two curves is related to the fact that the angle $\theta$
between the disorder-induced valley-Zeeman fields on the two dot
differs for the two disorder realizations.

We exploit the fact that the full Hamiltonian commutes with the axial ($z$)
component of the electron spin, and therefore one can identify four uncoupled
spin subspaces of the 22-dimensional two-electron Hilbert space.
It is beneficial to choose a classification corresponding to the standard two-electron
spin-singlet and spin-triplet states $\sket{S}$, $\sket{T_+}$, $\sket{T_0}$,
and $\sket{T_-}$ (the outer lower index refers to ``spin'').
Since these four spin subspaces are uncoupled from each other, 
the spin-Zeeman effect has no effect on the dynamics and therefore from now
on we disregard that.

In the absence of interdot tunneling, the energy eigenstates of the single-electron 
Hamiltonian are trivial: 
in the left dot they are 
$\ket{K_L \uparrow}$, $\ket{K_L \downarrow}$,
$\ket{K'_L \uparrow}$, and $\ket{K'_L \downarrow}$,
where 
$\ket{K_L}$ and $\ket{K'_L}$ are defined as the eigenstates of the 
$2\times 2$ matrix $\vec b_L^{\rm (tot)} \cdot \vec \tau$
corresponding to the eigenvalue $b_L^{\rm (tot)}$ and $-b_L^{\rm (tot)}$, respectively.
Single-electron energy eigenstates of the right dot are constructed 
accordingly.
The two-electron energy eigenstates are the standard Fock
basis states constructed from these single-electron states 
(as long as interdot tunneling is zero). 
The resulting 22 energy eigenstates are classified regarding their 
energy eigenvalue and spin state in Table \ref{tab:strongdisorder}.

\begin{table*}
\begin{ruledtabular}
\begin{tabular}{c|c|c|c|c}
  Energy &
    $\sket{S}$ &
    $\sket{T_+}$ &
    $\sket{T_0}$ &
    $\sket{T_-}$ \\
  \hline
  $b_L^{\rm (tot)} + b_R^{\rm (tot)}$ &
    $\frac{1}{\sqrt 2} \left( 
      \ket{K_L \uparrow,K_R \downarrow} - \ket{K_L \downarrow,K_R \uparrow}
       \right)$ &
    $\ket{K_L\uparrow,K_R\uparrow}$ &
        $\frac{1}{\sqrt 2} \left( 
      \ket{K_L \uparrow,K_R \downarrow} + \ket{K_L \downarrow,K_R \uparrow}
       \right)$ &
    $\ket{K_L\downarrow,K_R\downarrow}$
    \\
    $-b_L^{\rm (tot)} + b_R^{\rm (tot)}$ &
        $\frac{1}{\sqrt 2} \left( 
      \ket{K'_L \uparrow,K_R \downarrow} - \ket{K'_L \downarrow,K_R \uparrow}
       \right)$ &
    $\ket{K'_L\uparrow,K_R\uparrow}$ &
        $\frac{1}{\sqrt 2} \left( 
      \ket{K'_L \uparrow,K_R \downarrow} + \ket{K'_L \downarrow,K_R \uparrow}
       \right)$ &
    $\ket{K'_L\downarrow,K_R\downarrow}$
    \\
    $b_L^{\rm (tot)} - b_R^{\rm (tot)}$ &
        $\frac{1}{\sqrt 2} \left( 
      \ket{K_L \uparrow,K'_R \downarrow} - \ket{K_L \downarrow,K'_R \uparrow}
       \right)$ &
    $\ket{K_L\uparrow,K'_R\uparrow}$ &
        $\frac{1}{\sqrt 2} \left( 
      \ket{K_L \uparrow,K'_R \downarrow} + \ket{K_L \downarrow,K'_R \uparrow}
       \right)$ &
    $\ket{K_L\downarrow,K'_R\downarrow}$
    \\
    $-b_L^{\rm (tot)} - b_R^{\rm (tot)}$ &
    $\frac{1}{\sqrt 2} \left( 
      \ket{K'_L \uparrow,K'_R \downarrow} - \ket{K'_L \downarrow,K'_R \uparrow}
       \right)$ &
    $\ket{K'_L\uparrow,K'_R\uparrow}$ &
        $\frac{1}{\sqrt 2} \left( 
      \ket{K'_L \uparrow,K'_R \downarrow} + \ket{K'_L \downarrow,K'_R \uparrow}
       \right)$ &
    $\ket{K'_L\downarrow,K'_R\downarrow}$
    \\
  \hline
    $2b_R^{\rm (tot)}$ &
      $\ket{0,K_R \uparrow K_R \downarrow}$ &
    &
    &
    \\
    $0$ &
    $\frac{1}{\sqrt 2} \left( 
      \ket{0,K_R \uparrow K'_R \downarrow} - \ket{0,K_R \downarrow K'_R \uparrow}
       \right)$ &
    $\ket{0,K_R\uparrow K'_R\uparrow}$ &
        $\frac{1}{\sqrt 2} \left( 
      \ket{0,K_R \uparrow K'_R \downarrow} + \ket{0,K_R \downarrow K'_R \uparrow}
       \right)$ &
    $\ket{0,K_R\downarrow K'_R\downarrow}$
    \\
    $- 2b_R^{\rm (tot)}$ &
      $\ket{0,K'_R \uparrow K'_R \downarrow}$ &
    &
    &
\end{tabular}
\end{ruledtabular}
\caption{\label{tab:strongdisorder}
The 22 basis states used for perturbation calculations in the presence of strong
disorder. 
Different columns correspond to different spin states and each row has
a corresponding energy (left column; spin Zeeman energies are neglected). 
The upper four [lower three] rows contain the (1,1) states [(0,2) states].}
\end{table*}

At this point, we make use of the fact that the magnitudes of the total
valley-Zeeman-fields in the two dots are typically different (since they
have a random contribution induced by the random arrangement of disorder), 
and their difference is 
typically comparable to themselves
\bnen
\label{eq:cond}
b_L^{\rm (tot)} \sim b_R^{\rm (tot)} \sim b_L^{\rm (tot)}-b_R^{\rm (tot)}.
\eden
Note that this condition can hold only if the magnetic-field-induced 
valley-Zeeman field
$\Delta_v$ does not dominate over the disorder-induced component,
which restricts the validity of the following analysis to the range
$\Delta_v \lesssim b$.
If the condition \eqref{eq:cond} holds, then the separations between 
the seven energy levels considered in Table \ref{tab:strongdisorder}
are on the order of $b$, which is much larger than the interdot tunneling $t$, and 
therefore we are allowed to treat $t$
as a perturbation, and use the states listed in Table \ref{tab:strongdisorder} as the 
unperturbed states.

\begin{figure}
\includegraphics[scale=0.23]{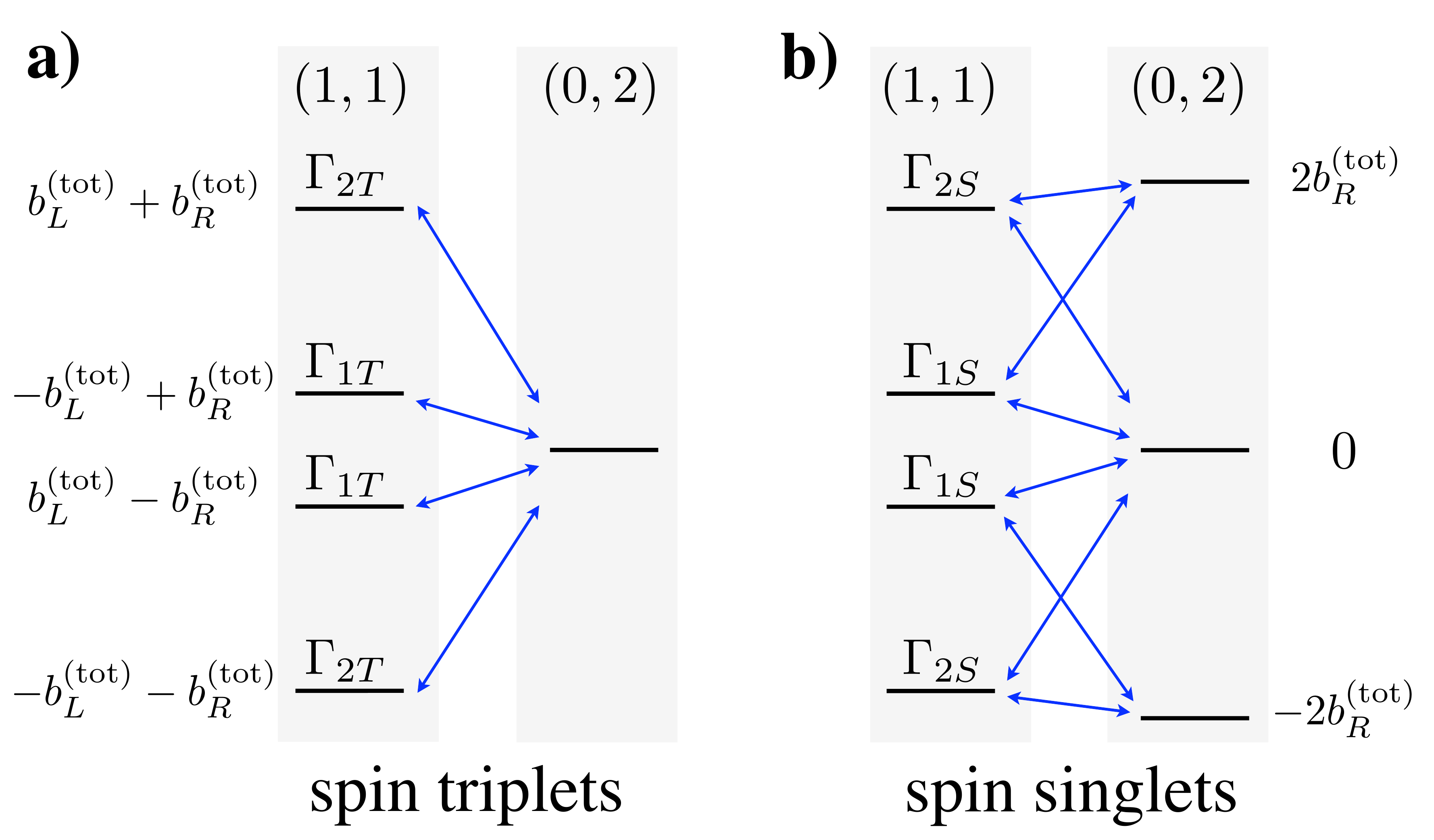}
\caption{\label{fig:levelsatstrongdisorder}
(Color online)
Two-electron states (lines) and their energies in a CNT DQD at 
strong disorder. Arrows denote tunnel couplings between (1,1) and (0,2) 
states.
Decay rates of (1,1) states are indicated, 
cf. Eqs. (\ref{eq:tripletrates} and \ref{eq:singletrates}) in text.
Here $b_R^{\rm (tot)} > b_L^{\rm (tot)}$.
(a) States and tunnel couplings in the five-dimensional spin-triplet subspaces.
The same plot refers to all three spin-triplet subspaces.
(b) States and tunnel couplings in the seven-dimensional spin-singlet subspace
(cf. Table \ref{tab:strongdisorder}).}
\end{figure}

Due to tunneling, the (1,1) states (upper four rows in Table \ref{tab:strongdisorder})
perturbatively hybridize with (0,2) states (lower three rows in Table  
\ref{tab:strongdisorder}) and therefore acquire a finite decay
rate. 
As shown in Table \ref{tab:strongdisorder} and Fig. \ref{fig:levelsatstrongdisorder}a,
in each of the three spin-triplet subspaces there are four (1,1) states and they 
hybridize with a single available (0,2) state.
Standard perturbation theory and Eq. \eqref{eq:decayrates}
gives two different decay rates,
\begin{subequations} \label{eq:tripletrates}
  \bean
    \frac{\Gamma_{1T}}{\Gamma_R} &=& 
      \frac{2t^2 \cos^2\left(\theta^{\rm (tot)}/2\right)}
      {\left(b_L^{\rm (tot)} - b_R^{\rm (tot)}\right)^2}, \\ 
    \frac{\Gamma_{2T}}{\Gamma_R} &=& 
      \frac{2t^2 \sin^2\left(\theta^{\rm (tot)}/2\right)}
      {\left(b_L^{\rm (tot)} + b_R^{\rm (tot)}\right)^2}.
  \eean
\end{subequations}
The rate $\Gamma_{1T}$ ($\Gamma_{2T}$) corresponds to the spin-triplet states in 
the second and third (first and fourth) lines of Table \ref{tab:strongdisorder}.
In the spin-singlet subspace, there are three (0,2) states to hybridize with, although
each (1,1) state hybridizes only with two (0,2) states 
(e.g., $\ket{K_L \uparrow ,K_R \downarrow} -  \ket{K_L \downarrow, K_R \uparrow}$
is not coupled to
$\ket{0,K'_R \uparrow,K'_R \downarrow}$ 
by tunneling). 
Due to hybridization, the four (1,1) spin-singlet states acquire two different decay rates,
\begin{subequations}
 \label{eq:singletrates}
 \bean 
   \frac{\Gamma_{1S}}{\Gamma_R} &=& 2t^2\left[
      \frac{\cos^2\left(\theta^{\rm (tot)}/2\right)}
      {\left(b_L^{\rm (tot)} - b_R^{\rm (tot)}\right)^2}+
      \frac{2\sin^2\left(\theta^{\rm (tot)}/2\right)} 
      {\left(b_L^{\rm (tot)} + b_R^{\rm (tot)}\right)^2}
    \right] , \nonumber \\ 
 \\
  \frac{\Gamma_{2S}}{\Gamma_R} &=& 2t^2\left[
      \frac{2\cos^2\left(\theta^{\rm (tot)}/2\right)}
      {\left(b_L^{\rm (tot)} - b_R^{\rm (tot)}\right)^2}+
      \frac{\sin^2\left(\theta^{\rm (tot)}/2\right)} 
      {\left(b_L^{\rm (tot)} + b_R^{\rm (tot)}\right)^2}
    \right]. \nonumber \\
 \eean
\end{subequations}
The rate $\Gamma_{1S}$ ($\Gamma_{2S}$) corresponds to the spin-singlet states in 
the second and third (first and fourth) lines of Table \ref{tab:strongdisorder}.
We emphasize that the valley dynamics of the spin-singlet subspace is remarkably 
different from 
the valley dynamics of the spin-triplet subspaces and 
the spin dynamics in the spin blockade of conventional semiconductor 
DQDs\cite{Jouravlev-spinblockade,Koppens-spinblockade}:
in the latter cases there is only a single available (0,2) state to hybridize with,
whereas in the former case there are three of them.

From now on we are aiming at deriving an analytical formula for the current in 
leading order in the small parameter $t/b$.
As the next step toward that we argue that the steady-state occupations of the
(0,2) states are negligible.
There are two facts needed to prove this.
(i) The steady-state current can be separated to contributions from single-electron 
tunneling via the (1,1) states and (0,2) states: 
$I = \sum_{\alpha \in (1,1)}\rho_{\alpha} \Gamma_{\alpha} 
+ \sum_{\alpha \in (0,2)}\rho_{\alpha}\Gamma_{\alpha}$.
For the (1,1) states, decay rates originate from a weak hybridization of the (0,2)
states, therefore in the first sum, 
$\Gamma_\alpha \sim (t/b)^2 \Gamma_R \ll \Gamma_R$.
For the (0,2) states, decay rates come from direct coupling to the right lead, hence
in the second sum, $\Gamma_\alpha \sim \Gamma_R$.
(ii) The (0,2) states are ``difficult to load'' and ``easy to empty'', 
therefore, as it can be
shown rigorously, their steady-state occupations are $\sim (t/b)^4$ whereas the 
occupations of the (1,1) states are $\sim 1$.
As a result concerning the current, this means that the contributions from the (1,1)
states provide the leading-order result, and the (0,2) states can be eliminated from
the classical master equation [Eq. \eqref{eq:2ecme}] .

The steady-state CME retrieved after the elimination can be solved
analytically using the ansatz 
\bnen
\label{eq:steadystate}
\rho_\alpha = 
  \frac{\Gamma_\alpha^{-1}}{\sum_{\alpha' \in (1,1)} \Gamma_{\alpha'}^{-1}},
  \ \ \ \ \ \  [\alpha \in (1,1)],
\eden
which expresses that the occupation probability of a state is proportional to the 
lifetime of that state.
The fact that this simple ansatz solves our classical master equation is a consequence
of the equivalence of the 16 (1,1) states in the following sense: 
if one of those is filled randomly with a uniform distribution, 
then after one transport cycle the occupations are still uniformly distributed.
Mathematically, the $16\times 16$ ``return probability matrix'' of the 
(1,1) states
\bnen
\label{eq:return}
R_{\alpha \beta} =
 \sum_{j \in (0,1)}
  \frac{r_{\alpha j}}{\sum_{\alpha'} r_{\alpha' j}}
  \frac{p_{j\beta}}{\sum_{j'} p_{j'\beta}},\ \ \ \ [\alpha,\beta \in (1,1)]
\eden
is doubly stochastic.
The key observation in proving this is that the row sums of 
the $(1,1)\to (0,1)$ transition probability matrix $p_{j\beta}/\sum_{j'} p_{j'\beta}$
are equal, which 
is a consequence of the vanishing detuning between (1,1) and (0,2) states.
Note that the connection between $R$ and the coefficient matrix $M$ of the CME is
\bnen
R_{\alpha \beta} = \left.\left(
\delta_{\alpha \beta} + \frac{M_{\alpha \beta}}{\sum_{k\in(0,1)} p_{k\beta}}
\right)\right|_{\alpha,\beta \in (1,1)}.
\eden

Using the solution in Eq. \eqref{eq:steadystate} 
and the rates listed in Eqs. \eqref{eq:tripletrates} and
\eqref{eq:singletrates},
we obtain an analytical formula from Eq. \eqref{eq:current}
for the steady-state current through the DQD,
\bnen
\label{eq:currentatstrongdisorder}
I = e \frac{16}
  {6\left(\Gamma^{-1}_{1T}+\Gamma^{-1}_{2T}\right) + 
  2\left(\Gamma^{-1}_{1S}+\Gamma^{-1}_{2S}\right)}.
\eden
This analytical result is compared to numerical results in Fig. \ref{fig:strongdisorder}
and a good correspondence is found.
Our analytical result gives an insight on how the angle $\theta^{\rm (tot)}$
influences the current: at angles close to $0$ and $\pi$, i.e., at parallel and 
antiparallel valley-Zeeman fields on the two dots, the current is suppressed,
since then either $\Gamma_{1T}$ or $\Gamma_{2T}$ is 
small and that makes the denominator in Eq. \eqref{eq:currentatstrongdisorder}
large. This qualitatively explains the zero-field dip in Fig. \ref{fig:strongdisorder} in
the case of $\theta=15\pi/16$:
as the magnetic field and hence $\Delta_v$ is increased, the angle $\theta^{\rm (tot)}$
crosses over from $15\pi/16$ toward $0$
(since the magnetic field is enforcing alignment of 
$\vec b_L^{\rm (tot)}$ and $\vec b_R^{\rm (tot)}$), 
starting from and ending at suppressed
current values, but sweeping through a region of enhanced current.

Our conclusion of this section is that in the considered regime 
the magnetotransport curve shows either a zero-field dip or peak,
depending on the disorder configuration.
Although our analysis in this section was based on the complete absence of 
spin-orbit coupling and detuning between (1,1) and (0,2) states, and
a perfect alignment of the magnetic field and the CNT axis, we expect no qualitative 
changes in the results in the case of weak spin-orbit interaction and detuning 
$\Dso,\Delta \ll b$ and/or a small misalignment of the field,
since those factors have no effect on the tunneling amplitude and can only 
slightly modify our unperturbed basis states and the corresponding energies.

As discussed in Sec. \ref{sec:strongsoi}, in
recent experiments\cite{Churchill-13cntprl,Churchill-cntspinblockade} 
with CNT DQDs a zero-field 
dip has been found with a dip width controllable by the tunneling amplitude.
In the analytical and numerical results of this section, the width of the dip
(predicted for certain disorder configurations) 
is insensitive to the tunneling amplitude because $t$ appears in the current as
a $t^2$ prefactor only.
Therefore, we conclude that the parameter regime of the measurement was probably
different from the one considered in this section.
This opinion is supported by the facts that in Ref.~ \onlinecite{Churchill-13cntprl} the 
ratio $\Dso/b \approx 7$ has been estimated and that
in Sec. \ref{sec:strongsoi},  in a different parameter regime
we have found a qualitative agreement with experiments.


\section{Conclusions}
\label{sec:discussion}

We have found that in the regime of strong
spin-orbit interaction, the magnetic-field dependence
of the leakage current shows a zero-field dip with a width tunable by
the interdot tunneling amplitude $t$ (Sec. \ref{sec:strongsoi}). 
We have shown that the ratio of the finite-field and the zero-field
current is typically $I_{\rm max}/I_0 \lesssim 1.5$.
Both the trend in the magnetotransport data (i.e., the zero-field dip) and the numerical
value $1.5$ agree well with those found in 
Ref.~ \onlinecite{Churchill-13cntprl} [see Fig. 3(a) therein].
In the measurement of Ref.~ \onlinecite{Churchill-cntspinblockade} [Fig. 3(e) therein]
 the qualitative
behavior is similar to our prediction, but a much larger ratio,
$I_{\rm max}/I_0\sim 50$ has been found, which deviates significantly from the 
prediction of our model.
This deviation might arise from the parameters of the measured sample
not fitting into the parameter regime considered in Section \ref{sec:strongsoi}.
(In Ref.~ \onlinecite{Churchill-cntspinblockade} the values of spin-orbit 
interaction energy and valley-mixing energy have not been
estimated. Interdot tunneling has been quoted as $t \sim 50 \mu$eV.)
Another potential reason for the deviation might be that certain
features and mechanisms possibly important in the spin-valley blockade are 
excluded from our model.
A relevant mechanism might be the spin- and/or valley-relaxation due to 
electron-phonon 
interaction or electron exchange with the leads\cite{Bulaev-socincntdots,Rudner-deflectioncoupling,Struck-spinrelaxation,
Danon-spinblockade,Vorontsov-spinblockade}.
If those relaxation rates are comparable to or larger than the lead-dot tunneling 
rates then they could affect the transport properties.
Another possibly influential effect disregarded in our model might
be the emergence of strongly correlated
Wigner-molecule-like states due to the strong electron-electron interaction
in nanotubes\cite{Roy-coulombincnt,Secchi-coulombincnt,Wunsch-coulombincnt},
which would imply the rearrangement of the energy level structure shown in Table 
\ref{tab:class} and Fig. \ref{fig:leveldiagram} and therefore could lead to a qualitatively 
different transport behavior.

We have studied the influence of disorder on the spin-valley blockade in the case of 
small spin-orbit interaction, where the dominant energy scale is that of the
short-range disorder (Sec. \ref{sec:strongdisorder}).
In this regime, the leakage current can show a zero-field dip or peak,
depending on the disorder configuration.
Although we are not aware of any measurements carried out in this regime, we think
that our results might be relevant for future experiments on graphene- and
silicon-based double quantum dots. 

Our QD model incorporating the valley-mixing effect due to disorder 
can serve as a starting point for future theoretical work on CNT QDs.
The fact that the valley-mixing effective Zeeman field depends on the 
electronic wave function and the disorder configuration felt by the electron
implies that this valley-Zeeman field changes as the electron is replaced.
This feature might allow for resonant electronic valley manipulation similar to recent
spin manipulation experiments in conventional semiconductor QDs
using spin-orbit coupling\cite{Nowack-esr} and
hyperfine interaction\cite{Laird-prl-edsr}.
Note that in clean nanotubes, such resonant techniques do not require a 
magnetic field since the two valley states having the same spin 
are split by the spin-orbit splitting even at zero field.
A further possible application of our model could be to describe the pulsed-gate
experiments of Churchill \emph{et al.}\cite{Churchill-13cntprl} which intended
to infer relaxation and decoherence times of two-electron spin-valley states in a 
nanotube double dot.
As those results have been obtained using isotope-enriched samples, incorporating
the spin- and valley-mixing hyperfine interaction\cite{Palyi-spinblockade} might 
also be necessary.

In conclusion, we have found that spin-independent 
short-range disorder in carbon nanotube
double quantum dots can lift the spin-valley blockade.
In our transport model, we account for valley degeneracy, spin-orbit energy splitting, 
and disorder-induced valley mixing, which are characteristic features of nanotube
quantum dots distinguishing them from their counterparts in 
conventional semiconductors.
The main result of this work is that in the regime of strong spin-orbit interaction
our model predicts a zero-field dip in the magnetic-field dependence of the 
leakage current, with a dip width tunable by the height of the interdot tunneling barrier.
This behavior is in accordance with recent experiments. 
Our analysis for the regime of strong disorder, which has possible relevance
for graphene- and silicon-based double quantum dots, predicts that the 
magnetotransport shows either a zero-field dip or peak depending on the 
disorder configuration. 

{\it Note added in proof.} After completion and submission of this work we became
aware of two related theoretical studies,\cite{vonStecher,Weiss} which
focus on spectral properties of perfectly clean (disorder-free) CNT DQDs.
The work of von Stecher {\it et al.}\cite{vonStecher} also describes the effect
of strong correlations on transport in the Pauli blockade regime. 
The purpose of the present work is to account for disorder-induced effects
in the spin-valley blockade, which makes it distinct from Refs.\,
\onlinecite{vonStecher} and \ \onlinecite{Weiss}.

\begin{acknowledgments}
We thank DFG for financial support within Grants No. SFB 767, No. SPP 1285, and
No. FOR 912.
\end{acknowledgments}


\appendix

\section{Strong spin-orbit coupling and $\frac{b_-^2}{\Dso} \gg t$}
\label{app:strongdisorder}

We have shown in Sec. \ref{sec:strongsoi} that in the regime of 
strong spin-orbit interaction and $\frac{b_-^2}{\Dso} \sim t$,
the magnetotransport curve develops a zero-field dip with a width which is 
controllable by the tunneling amplitude $t$.
Here we argue that the above condition can be relaxed and the
same statement is true in the regime $\frac{b_-^2}{\Dso} \gg t$
(provided $\Dso \gg b,t,\Delta_v$).

We recall that in Sec. \ref{sec:strongsoi} we have found that at zero field six out of
the 22 two-electron states are blocked, and two out of those six show a 
blocked-unblocked crossover as the magnetic field is switched on.
We have determined the field-dependent decay rates of these six states without 
using the assumption $\frac{b_-^2}{\Dso} \sim t$, therefore those results hold in
the current case $\frac{b_-^2}{\Dso} \gg t$ as well.

The effective Hamiltonians in Eqs. \eqref{eq:h0pm} and \eqref{eq:hpm0} are
also valid in the current case.
However, from this point there is an important deviation in the procedure
compared to Sec. \ref{sec:strongsoi}.
In the zero-energy spin-polarized subspaces [described by $H_{0,\pm}$
in Eq. \eqref{eq:h0pm}], the three basis state does not mix evenly but instead the
(1,1) supertriplet and supersinglet states hybridize strongly with each other
and these hybridized states themselves
hybridize only weakly with the (0,2) supersinglet via tunneling.
This leads to decay rates 
$\sim\frac{\Dso^2 t^2}{b_-^4} \Gamma_R \ll \Gamma_R$
(in contrast to Sec. \ref{sec:strongsoi} where these rates were found to be
$\sim \Gamma_R$).
An important point is that if $b_a \sim b_-$,which is typically true due to the random 
nature of the disorder-induced valley fields $\vec b_L$ and $\vec b_R$, 
then the decay rate $\sim \frac{\Dso^2 t^2}{b_-^4} \Gamma_R$
is still orders-of-magnitude larger than the decay rate $\Gamma_s$ 
of the blocked states, and therefore these states can still be considered as unblocked.
The same argument applies for the six states in the zero-energy mixed-spin subspace,
therefore each of those can be classified as unblocked.

From these results we conclude that in the case $\frac{b_-^2}{\Dso} \gg t$
the set of blocked states (which determine the character of the magnetotransport
curve) and the form of their decay rates are the same as found in 
the regime $\frac{b_-^2}{\Dso} \sim t$, and hence the conclusions drawn there
hold here as well. This finding is confirmed by numerical calculations (not shown).

\section{Strong spin-orbit coupling and $\frac{b_-^2}{\Dso} \ll t$}
\label{app:strongdisorder2}

Here we argue that the zero-field magnetotransport dip, discussed in 
Sec. \ref{sec:strongsoi} and Appendix \ref{app:strongdisorder}, 
gets smeared out if the tunneling amplitude is increased to
the regime $\frac{b_-^2}{\Dso} \ll t$ such that
$t\gg b_-\sqrt{\Dso/b_a}$. 
(The latter condition is stronger than the former one provided that 
$\vec b_L \neq \vec b_R$ and therefore $b_a \neq 0$.)
As before, we restrict the discussion to the strong spin-orbit coupling regime: 
$\Dso \gg b,t,\Delta_v$.

As pointed out in the analysis after Eq. \eqref{eq:hpm0},
the blocked-unblocked crossover of the high- and low energy mixed-spin
supertriplets, which gives rise to a zero-field dip in the magnetotransport, 
occurs around the magnetic field where
$\Delta_{v} = \frac{t^2 b_a}{\sqrt{2} b_-^2}$.
This implies that if $t\gg b_-\sqrt{\Dso/b_a}$, then the crossover would
take place only in the high-field regime $\Delta_v \gg \Dso$ and not 
in the low-field regime $\Delta_v \ll \Dso$ under consideration.
As a consequence, the character of the 
low-field magnetotransport curve will be determined by the
field-induced evolution of the slow decay rates,
resulting in a parabolic peak around zero field.
This finding is confirmed by numerical calculations (not shown).


\bibliography{cnt-spinblockade}

\begin{thebibliography}{59}
\bibitem{lowdinfootnote} 
In the original work of L\"owdin (Ref.~ \onlinecite{Lowdin-partitioning}), 
the effective Hamiltonians derived for 
degenerate subspaces depend on the energy itself, leading to equations to be
solved self-consistently. 
In this work, we always 
approximate the energy in the effective Hamiltonian as
the unperturbed energy of the considered subspace, which results in an
energy-independent effective Hamiltonian.
\end{thebibliography}

\end{document}